\documentclass[showpacs,showkeys,preprintnumbers,amssymb,amsmath,aps,superscriptaddress,prb,twocolumn]{revtex4}
\usepackage{graphicx}
\usepackage{dcolumn}
\usepackage{bm}
\usepackage{epsfig}
\usepackage{color}

\begin{document}


\title{Exact ground states of a spin-1/2 Ising-Heisenberg model on the Shastry-Sutherland lattice in a magnetic field}

\author{Taras Verkholyak}
\email{werch@icmp.lviv.ua}
\affiliation{Institute for Condensed Matter Physics, NASU, 1 Svientsitskii Street, L'viv-11, 79011, Ukraine}
\affiliation{Institute of Physics, Faculty of Science, P. J. \v{S}af\'{a}rik University, Park Angelinum 9, 040 01, Ko\v{s}ice, Slovakia}
\author{Jozef Stre\v{c}ka}
\affiliation{Institute of Physics, Faculty of Science, P. J. \v{S}af\'{a}rik University, Park Angelinum 9, 040 01, Ko\v{s}ice, Slovakia}
\author{Fr\'ed\'eric Mila}
\affiliation{Institute of Theoretical Physics, Ecole Polytechnique F\'ed\'erale de Lausanne (EPFL), 1015 Lausanne, Switzerland}
\author{Kai P. Schmidt}
\affiliation{Lehrstuhl f\"ur Theoretische Physik I, Otto-Hahn-Strasse 4, TU Dortmund, 44221 Dortmund, Germany}

\date{\today}

\begin{abstract}
Exact ground states of a spin-1/2 Ising-Heisenberg model on the Shastry-Sutherland lattice with Heisenberg intra-dimer and Ising inter-dimer couplings are found by two independent rigorous procedures. The first method uses a unitary transformation to establish a mapping correspondence with an effective classical spin model, while the second method relies on the derivation of an effective hard-core boson model by continuous unitary transformations. Both methods lead to equivalent effective Hamiltonians providing a convincing proof that the spin-1/2 Ising-Heisenberg model on the Shastry-Sutherland lattice exhibits a zero-temperature magnetization curve with just two intermediate plateaus at one-third and one-half of the saturation magnetization, which correspond to stripe and checkerboard orderings of singlets and polarized triplets, respectively. The nature of the remarkable stripe order relevant to the one-third plateau is thoroughly investigated with the help of the corresponding exact eigenvector. The rigorous results for the spin-1/2 Ising-Heisenberg model on the Shastry-Sutherland lattice are compared with the analogous results for the purely classical Ising and fully quantum Heisenberg models. 
Finally, we discuss to what extent the critical fields of SrCu$_2$(BO$_3$)$_2$ and (CuCl)Ca$_2$Nb$_3$O$_{10}$ can be described within the suggested Ising-Heisenberg model.
\end{abstract}


\maketitle

\section{Introduction} 

The spin-1/2 quantum Heisenberg model on a two-dimensional orthogonal-dimer lattice has attracted considerable attention since the pioneering work by Shastry and Sutherland, which has rigorously proved that the ground state is exactly dimerized provided the inter-dimer coupling is not stronger than a half of the intra-dimer coupling.\cite{shas81} Later on, it has been verified by numerous analytical and numerical methods that the singlet-dimer state remains the true ground state even in a wider parameter range, which is limited just by the upper value $J'/J \approx 0.675$ of the interaction ratio between the inter-dimer and intra-dimer couplings.\cite{miya99,weih99,mull00,koga00,weih02,miya03,jlou12,corb13} A lot of efforts have been subsequently devoted to the magnetization process of this frustrated quantum spin model, which additionally reveals several intriguing quantum ground states that macroscopically manifest themselves as intermediate magnetization plateaus.\cite{momo00,misg01,aben08,meng08,dori08,isae09,taki13,corb14} Despite considerable efforts, there is still controversy and intense debate about the total number, size and microscopic nature of some intermediate magnetization plateaus.  

Almost two decades after the spin-1/2 quantum Heisenberg model on the Shastry-Sutherland lattice was originally invented, the first experimental realization of this rather curious theoretical model has been found in the layered copper-based compound SrCu$_2$(BO$_3$)$_2$. The magnetic compound SrCu$_2$(BO$_3$)$_2$ has thus offered a long sought experimental verification of the singlet-dimer state theoretically predicted by Shastry and Sutherland,\cite{shas81} because the actual ratio between the inter-dimer and intra-dimer couplings $J'/J \approx 0.63$ is sufficiently small in order to fall into the parameter range where the product of singlet dimers is the exact ground state. Early high-field magnetization measurements for SrCu$_2$(BO$_3$)$_2$ have come up with convincing evidence of three sizable plateaus at 1/8, 1/4 and 1/3 of the full magnetization in addition to the expected plateau at zero magnetization that corresponds to the singlet-dimer state.\cite{kage99,oniz00,kage01} Subsequent torque measurements performed by Sebastian et al.\cite{seba08} suggested the presence of several additional plateaus besides the three most sizable plateaus mentioned previously. Steady field experiments supported by NMR results\cite{taki13} have established the low-magnetization sequence of plateaus to be 1/8, 2/15, 1/6 and 1/4. At very high field, the first report in favor of a 1/2 plateau\cite{seba08,jaime12} has been confirmed by recent magnetization data for SrCu$_2$(BO$_3$)$_2$ recorded at ultrahigh magnetic fields which definitely established the presence of a robust magnetization plateau at 1/2 of the saturation magnetization, the width of which is nearly a half of that recorded for the most extensive 1/3 plateau.\cite{mats13}     

Another excellent realization of a magnetic structure relevant to the Shastry-Sutherland lattice is provided by a rather extensive class of isostructural rare-earth tetraborides RB$_4$ (R = Dy, Er, Tm, Tb, Ho).\cite{wata05,mich06,yosh06,yosh08,siem08,gaba08,mata10,jkim10,mats11} However, the magnetic behavior of the rare-earth tetraborides RB$_4$ is basically affected by the Ising (easy-axis) anisotropy due to strong crystal-field effects acting on rare-earth ions in contrast to the almost isotropic magnetic behavior of the transition-metal copper ions in SrCu$_2$(BO$_3$)$_2$. The metallic character along with the substantial Ising anisotropy make a comprehensive description of magnetic properties of the rare-earth tetraborides much more complex, because one has to take into account the coupling between spin and electronic subsystems described in terms of Ising (or $XXZ$ Heisenberg) and Falicov-Kimball models on the Shastry-Sutherland lattice, respectively.\cite{fark10,fark14} 

In the present work, two independent rigorous analytical methods will be employed for investigating the ground state of the spin-1/2 Ising-Heisenberg model on the Shastry-Sutherland lattice in a magnetic field, which accounts for the $XXZ$ Heisenberg intra-dimer and Ising inter-dimer couplings. The main goal for our study is to identify the microscopic nature of spin arrangements emerging within intermediate magnetization plateaus through exact eigenstates of the spin-1/2 Ising-Heisenberg model on the Shastry-Sutherland lattice, which may also have interesting implications for the magnetization plateaus experimentally observed for SrCu$_2$(BO$_3$)$_2$ and RB$_4$, as well as exact eigenstates of the full quantum Heisenberg counterpart model.  

The outline of this paper is as follows. In Sec.~\ref{sec-model} the Ising-Heisenberg model on the Shastry-Sutherland lattice is defined and the basic steps of its rigorous treatment are explained.
The most interesting results for the ground-state phase diagram and the nature of the spin arrangements emerging in intermediate magnetization plateaus are discussed in Sec.~\ref{sec-gs}. 
Finally, the most important outcomes of our work are briefly summarized in Sec.~\ref{sec-conclusions}.

\section{The Ising-Heisenberg model}
\label{sec-model}

\begin{figure*}
\begin{center}
\epsfig{file=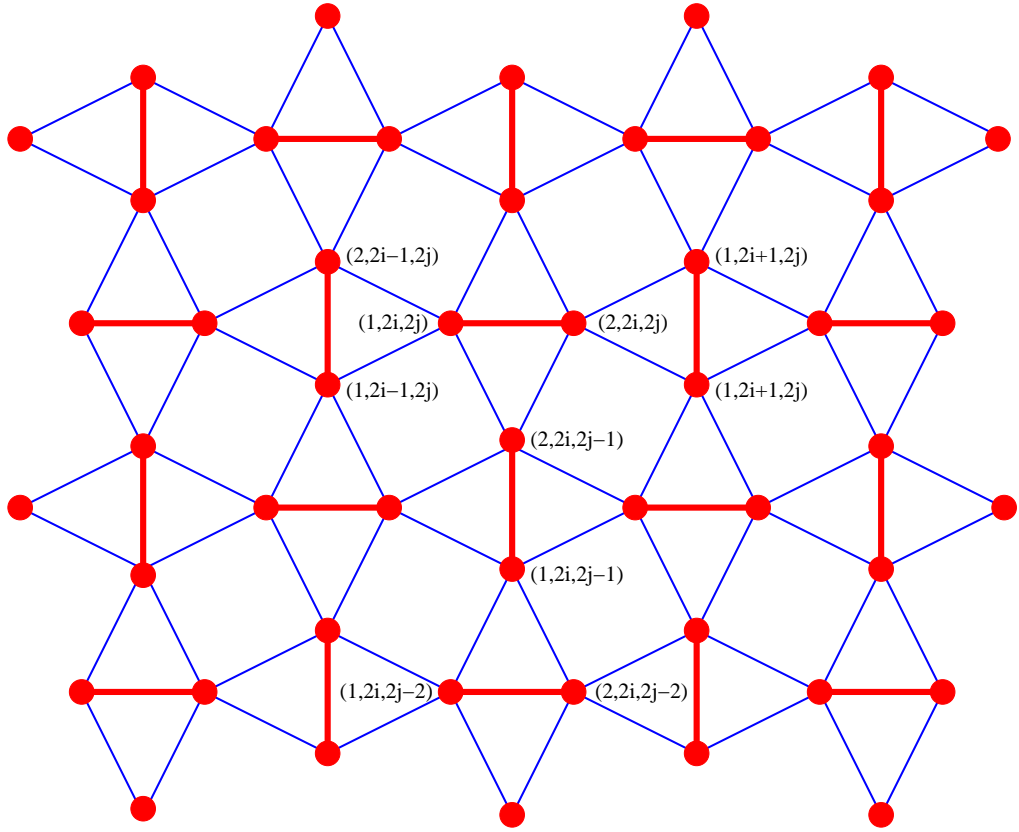, width=0.75\textwidth}
\end{center}
\caption{(Color online) Schematic illustration of the spin-1/2 Ising-Heisenberg model on the Shastry-Sutherland lattice with the $XXZ$ Heisenberg intra-dimer interaction $J (\Delta)$ and the Ising inter-dimer interaction $J'$. Thin (blue) lines show the Ising coupling and thick (red) lines denote the Heisenberg coupling. Spins inside the Heisenberg dimers are enumerated from left to right
and from bottom to top, respectively (see also Fig.~\ref{fig_node}).}
\label{fig_ssl}
\end{figure*}

Let us consider the spin-1/2 Ising-Heisenberg model on the Shastry-Sutherland lattice with $XXZ$ Heisenberg intra-dimer interaction $J (\Delta)$ and Ising inter-dimer interaction $J'$ defined through the following Hamiltonian: 
\begin{eqnarray}
H&=& J \sum_{i,j=1}^N ({\mathbf s}_{1,i,j}\cdot{\mathbf s}_{2,i,j})_\Delta 
  - h \sum_{i,j=1}^N (s_{1,i,j}^z{+}s_{2,i,j}^z)
\nonumber \\
  &+& J'\sideset{}{'}\sum_{i,j=1}^N (s_{1,i,j}^z{+}s_{2,i,j}^z)(s_{1,i+1,j}^z{+}s_{2,i-1,j}^z) 
\nonumber \\
 &+& J'\sideset{}{''}\sum_{i,j=1}^N (s_{1,i,j}^z{+}s_{2,i,j}^z)(s_{1,i,j+1}^z{+}s_{2,i,j-1}^z),
\label{ham-def}
\end{eqnarray}
where $({\mathbf s}_{1,i,j}\cdot{\mathbf s}_{2,i,j})_\Delta =\Delta(s_{1,i,j}^x s_{2,i,j}^x + s_{1,i,j}^y s_{2,i,j}^y) + s_{1,i,j}^z s_{2,i,j}^z$, $s_{l,i,j}^\alpha$ denotes spatial projections ($\alpha=x,y,z$) of the spin-$1/2$ operator, the first index $l=1,2$ enumerates the spins inside of the Heisenberg dimer, the second and third index determines the position of the dimer on a virtual square lattice by specifying its column and row, respectively (see Fig.~\ref{fig_ssl}). The first and second summations are carried out over all dimers in order to account for the anisotropic $XXZ$ Heisenberg intra-dimer interaction $J$($\Delta$) and  the Zeeman's magnetostatic energy of the spins in an external magnetic field $h$,
while the third (fourth) summation $\sum'$ ($\sum''$) restricted by the constraint $i+j=odd$ ($i+j=even$) extends over all vertical (horizontal) dimers to account for the Ising inter-dimer interaction $J'$. 
\begin{figure*}
\epsfig{file=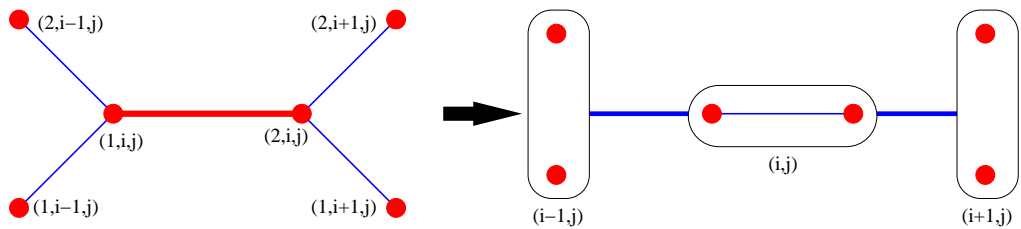, width=0.75\textwidth}
\caption{(Color online) Schematic representation of the local cluster Hamiltonian $H_{[i-1:i+1],j}$ formed by three consecutive dimers in the horizontal direction and of its reduction to the Ising-like form.}
\label{fig_node}
\end{figure*}

The model defined by the Hamiltonian (\ref{ham-def}) can be alternatively viewed as an assembly of spin-1/2 Heisenberg dimers on a fictitious square lattice composed of two interpenetrating sublattices: one sublattice of horizontal dimers ($i+j=even$) and one of vertical dimers ($i+j=odd$). The nearest-neighbor spins from the dimers belonging to different sublattices are coupled by the Ising inter-dimer interaction. For further convenience, it is useful to rewrite the total Hamiltonian (\ref{ham-def}) as a sum of local cluster Hamiltonians:

\begin{equation}
\label{ham0}
H=
\sideset{}{'}\sum_{i,j=1}^N H_{i,[j-1:j+1]}
+\sideset{}{''}\sum_{i,j=1}^N H_{[i-1:i+1],j},
\end{equation}
\begin{eqnarray}
\label{ham-local0}
&&H_{[i-1:i+1],j}{=}
J({\mathbf s}_{1,i,j}\cdot{\mathbf s}_{2,i,j})_\Delta
{-}h(s_{1,i,j}^z{+}s_{2,i,j}^z)
\\
&&
{+}J'[(s_{1,i-1,j}^z{+}s_{2,i-1,j}^z)s_{1,i,j}^z{+}s_{2,i,j}^z(s_{1,i+1,j}^z{+}s_{2,i+1,j}^z)],
\nonumber\\
&&H_{i,[j-1:j+1]}{=}
J({\mathbf s}_{1,i,j}\cdot{\mathbf s}_{2,i,j})_\Delta
{-}h(s_{1,i,j}^z{+}s_{2,i,j}^z)
\nonumber\\
&&
{+}J'[(s_{1,i,j-1}^z{+}s_{2,i,j-1}^z)s_{1,i,j}^z{+}s_{2,i,j}^z(s_{1,i,j+1}^z{+}s_{2,i,j+1}^z)],
\nonumber
\end{eqnarray}
which include all interaction terms between the nearest-neighboring spins from spin clusters constituted by three consecutive dimers arranged either in a horizontal or vertical direction (see the spin cluster on the left-hand-side of Fig. \ref{fig_node}). Owing to the specific form of the Hamiltonian, the $z$-component of the total spin $S_{i,j}^z = s_{1,i,j}^z+s_{2,i,j}^z$ of each Heisenberg dimer commutes with the total Hamiltonian (\ref{ham0}) as well as with each local cluster Hamiltonian (\ref{ham-local0}). Hence, it follows that the $z$-component of the total spin $S_{i,j}^z$ is a conserved quantity with well defined quantum spin numbers and, consequently, all local cluster Hamiltonians (\ref{ham-local0}) also commute with each other. This property is of fundamental importance for the reduction of the total Hamiltonian (\ref{ham0}) into a diagonal (Ising-like) representation, which can be performed by two independent approaches either based on local or continuous unitary transformations. 

\subsection{Local unitary transformations}

\begin{figure*}[tbh]
\epsfig{file=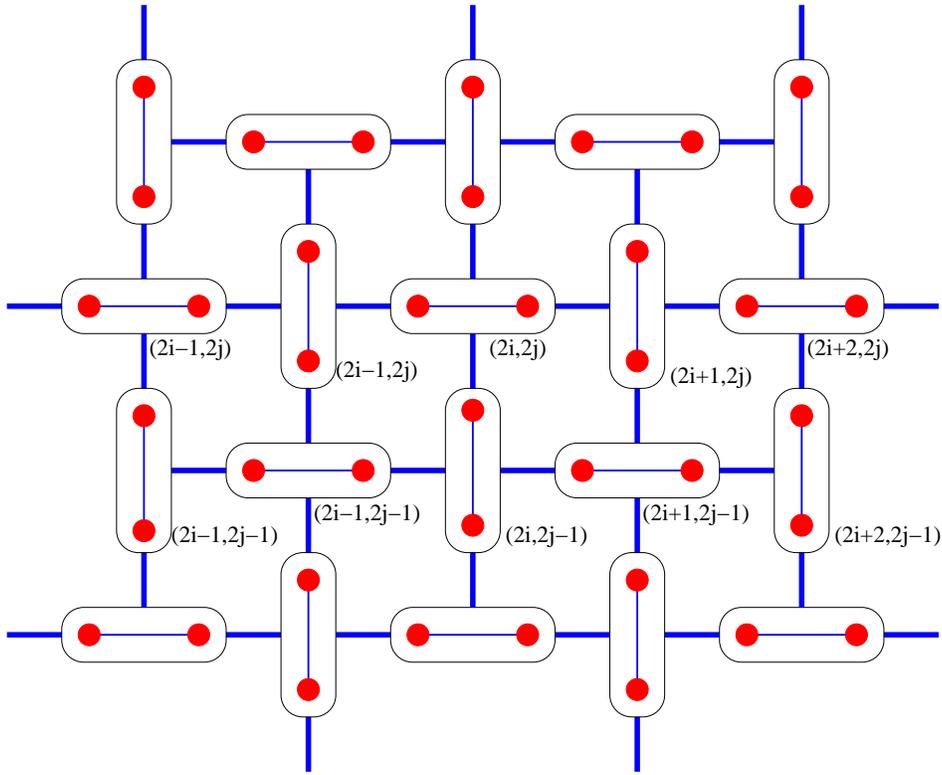, width=0.7\textwidth}
\caption{(Color online) Schematic representation of the effective classical spin model on the Shastry-Sutherland lattice obtained by applying the local unitary transformation (\ref{U}) to each Heisenberg dimer.}
\label{fig_ssl-dim}
\end{figure*}

At first, let us briefly describe the basic steps of the first method based on the local unitary transformation for the spin-1/2 Heisenberg dimers. It is worthy to notice that the local cluster Hamiltonians (\ref{ham-local0}) are already diagonal in a particular subspace $S_{i,j}^z = \pm 1$ of the Heisenberg dimers with equally oriented spins. To diagonalize the local cluster Hamiltonians (\ref{ham-local0}) in the other subspace $S_{i,j}^z = 0$ spanned by two oppositely oriented spins of the Heisenberg dimers, one may use the local unitary transformation acting nontrivially in this subspace only:
\begin{eqnarray}
\label{U}
U_{i,j}&=&\left(\frac{1}{2}+2s_{1,i,j}^z s_{2,i,j}^z\right)
\nonumber\\
&+&\exp{[i2\alpha_{i,j}s_{1,i,j}^x s_{2,i,j}^y]}\left(\frac{1}{2}-2s_{1,i,j}^z s_{2,i,j}^z\right),
\end{eqnarray}
where the parameter $\alpha_{i,j}$ is defined as follows:
\begin{widetext}
\begin{eqnarray}
\cos\alpha_{i,j} &=&\frac{J'(S_{i+1,j}^z-S_{i-1,j}^z)}{\sqrt{\Delta^2J^2 + J'^2(S_{i+1,j}^z-S_{i-1,j}^z)^2 }}, \;
\sin\alpha_{i,j}=\frac{\Delta J}{\sqrt{\Delta^2J^2 + J'^2(S_{i+1,j}^z-S_{i-1,j}^z)^2 }}, \;
\mbox{for} \: i+j=even,
\nonumber\\
\cos\alpha_{i,j}&=&\frac{J'(S_{i,j+1}^z-S_{i,j-1}^z)}{\sqrt{\Delta^2J^2 + J'^2(S_{i,j+1}^z-S_{i,j-1}^z)^2 }}, \;
\sin\alpha_{i,j}=\frac{\Delta J}{\sqrt{\Delta^2J^2 + J'^2(S_{i,j+1}^z-S_{i,j-1}^z)^2 }},\;
\mbox{for} \: i+j=odd.
\label{tp}
\end{eqnarray}
\end{widetext}
It is quite evident from Eq.~(\ref{tp}) that the transformation parameter $\alpha_{i,j}$ for the horizontal (vertical) Heisenberg dimer depends on the $z$-component of the total spin on two adjacent vertical (horizontal) Heisenberg dimers. Applying the unitary transformation (\ref{U}) to the local cluster Hamiltonian (\ref{ham-local0}) one obtains the following diagonal (Ising-like) representation of the local cluster Hamiltonians:
\begin{widetext}
\begin{eqnarray}
\label{ham-local1}
 U_{i,j}H_{[i-1:i+1],j}U_{i,j}^{+}
&=&\frac{|\Delta J|}{2}(s_{2,i,j}^z-s_{1,i,j}^z) + Js_{1,i,j}^z s_{2,i,j}^z 
-h(s_{1,i,j}^z+s_{2,i,j}^z)
+\frac{1}{2}(s_{2,i,j}^z-s_{1,i,j}^z)I(|S_{i+1,j}^z-S_{i-1,j}^z|)
\nonumber\\
&&+\frac{J'}{2}(s_{1,i,j}^z+s_{2,i,j}^z)(S_{i+1,j}^z+S_{i-1,j}^z),
\nonumber\\
U_{i,j}H_{i,[j-1:j+1]}U_{i,j}^{+}
&=&\frac{|\Delta J|}{2}(s_{2,i,j}^z-s_{1,i,j}^z) + Js_{1,i,j}^z s_{2,i,j}^z 
-h(s_{1,i,j}^z+s_{2,i,j}^z)
+\frac{1}{2}(s_{2,i,j}^z-s_{1,i,j}^z)I(|S_{i,j+1}^z-S_{i,j-1}^z|)
\nonumber\\
&&+\frac{J'}{2}(s_{1,i,j}^z+s_{2,i,j}^z)(S_{i,j+1}^z+S_{i,j-1}^z),
\end{eqnarray}
where
\begin{eqnarray}
\label{eff_interaction}
&&\!\!\!\!\!\!\!\!\!\!\!\!
I(|S_{i+1,j}^z{-}S_{i-1,j}^z|){=}\delta(|S_{i+1,j}^z{-}S_{i-1,j}^z|{-}1)\left(\sqrt{\Delta^2 J^2{+}J'^2}{-}|\Delta J|\right)
{+}\delta(|S_{i+1,j}^z{-}S_{i-1,j}^z|{-}2)\left(\sqrt{\Delta^2 J^2{+}4J'^2}{-}|\Delta J|\right)\geq 0,
\\
\label{projections}
&&\!\!\!\!\!\!\!\!\!\!\!\!
\delta(|S_{i+1,j}^z-S_{i-1,j}^z|-1)=[(S_{i+1,j}^z)^2-(S_{i-1,j}^z)^2]^2, \;
\delta(|S_{i+1,j}^z-S_{i-1,j}^z|-2)=\frac{1}{2}S_{i-1,j}^zS_{i+1,j}^z(S_{i-1,j}^zS_{i+1,j}^z-1).
\end{eqnarray}
\end{widetext}
Here, the symbol $\delta(\dots)$ is used for the Kronecker delta function. It can be readily understood from Eq.~(\ref{ham-local1}) that the transverse $XX$-part of the Heisenberg intra-dimer coupling produces due to the local unitary transformation (\ref{U}) an effective staggered field of magnitude $\Delta J/2$ and a more complex effective multispin interaction, whose specific form depends basically mainly on the difference between the $z$-components of the total spin on two adjacent Heisenberg dimers. A graphical representation of the unitary transformation (\ref{U}) is depicted in Fig.~\ref{fig_node}. Another important implications follow from the commutation relation $[U_{i,j},H_{[i'-1:i'+1],j'}]=[U_{i,j},H_{i,[j'-1:j'+1]}]=0$ for $i\neq i'$ or $j\neq j'$. Owing to this fact, one may separately apply the unitary transformation (\ref{U}) to each Heisenberg dimer and consequently, the whole Hamiltonian (\ref{ham0}) can be reduced to the following Ising-like (diagonal) representation:
\begin{eqnarray}
\tilde{H}=\sum_{i,j=1}^{N} \tilde{H}_{i,j}^{0}
+\sideset{}{'}\sum_{i,j=1}^N \tilde{V}_{i,[j-1:j+1]}
+ \sideset{}{''}\sum_{i,j=1}^N \tilde{V}_{[i-1:i+1],j},
\label{h-2d-od}
\end{eqnarray}
where
\begin{widetext}
\begin{eqnarray}
&&\!\!\!\!\!\!\!\!
\tilde{H}_{i,j}^0=
\frac{|\Delta J|}{2}(s_{2,i,j}^z-s_{1,i,j}^z) + Js_{1,i,j}^z s_{2,i,j}^z
 - h (s_{1,i,j}^z + s_{2,i,j}^z),
\nonumber\\
&&\!\!\!\!\!\!\!\!
\tilde{V}_{[i{-}1:i{+}1],j}{=}
\frac{1}{2}(s_{2,i,j}^z{-}s_{1,i,j}^z) I(|s_{1,i{+}1,j}^z {+} s_{2,i{+}1,j}^z{-}s_{1,i{-}1,j}^z{-}s_{2,i{-}1,j}^z|)
{+}\frac{J'}{2}(s_{1,i,j}^z {+} s_{2,i,j}^z)(s_{1,i{+}1,j}^z {+} s_{2,i{+}1,j}^z{+}s_{1,i{-}1,j}^z{+}s_{2,i{-}1,j}^z)
\nonumber\\
&&\!\!\!\!\!\!\!\!
\tilde{V}_{i,[j{-}1:j{+}1]}{=}
\frac{1}{2}(s_{2,i,j}^z{-}s_{1,i,j}^z) 
I(|s_{1,i,j{+}1}^z{+}s_{2,i,j{+}1}^z{-}s_{1,i,j{-}1}^z{-}s_{2,i,j{-}1}^z|)
{+}\frac{J'}{2}(s_{1,i,j}^z {+} s_{2,i,j}^z)
(s_{1,i,j{+}1}^z {+} s_{2,i,j{+}1}^z{+}s_{1,i,j{-}1}^z{+}s_{2,i,j{-}1}^z).
\nonumber\\
\label{h-2d-odd}
\end{eqnarray}
\end{widetext}
The schematic representation of the classical spin model defined by the effective Hamiltonian (\ref{h-2d-od}) is presented in Fig.~\ref{fig_ssl-dim}.

\subsection{Continuous unitary transformations}

Next we are aiming at an alternative derivation of an effective low-energy model based on continuous unitary transformations which also gives the magnetization curve of the Ising-Heisenberg model. In its perturbative formulation \cite{knet00,knet03} which can be applied for $\Delta=1$, this method has been already applied succesfully for the full quantum Heisenberg model on the two-dimensional Shastry-Sutherland lattice \cite{dori08} as well as on quasi one-dimensional variants of the lattice \cite{manm11,folt14}.
Here we apply also the method of perturbative continuous unitary transformations (pCUTs) along the same lines, but more importantly, we show that the recently formulated non-perturbative graph-based continuous unitary transformations (gCUTs) \cite{yang11} yield the exact low-energy model for any $\Delta$ in agreement with the approach of the previous section based on local unitary transformations.

First, we rewrite the Ising-Heisenberg model in the form
\begin{eqnarray}
\label{eqn:Hpcut}
\frac{\hat H}{J} &=& \sum_{\langle i,j \rangle} \vec{S}_i \cdot \vec{S}_j  + x\sum_{\langle\langle i,j \rangle\rangle } S_i^z\,  S_j^z + \frac{h}{J} \sum_i S_i^z\nonumber\\
       &=& {-}\left(\frac{3}{4}\right)^{N_{\rm d}}{+}\sum_{\nu,\alpha}\hat{t}^{\dagger}_{\nu,\alpha}\hat{t}^{\phantom{\dagger}}_{\nu,\alpha} {+}x \left( \hat{T}_{-1}{+}\hat{T}_0{+}\hat{T}_{1}\right) {+} \frac{h}{J} \hat{H}_{h}\nonumber\\
       &=& E_0 + \hat{Q} +x \left( \hat{T}_{-1}+\hat{T}_0+\hat{T}_{1}\right) +\frac{h}{J} \hat{H}_{h}\quad ,
\end{eqnarray}
where we use different notation for the sake of convenience, i.e. $i$, $j$ enumerate the spins, 
and $\langle i,j \rangle$ ($\langle\langle i,j \rangle\rangle$) stands for the summation over all intra-dimer (inter-dimer) interaction. Finally, $x=J'/J$ corresponds to the natural perturbation parameter inside the singlet-dimer phase and $\hat{t}^{\dagger}_{\nu,\alpha}$ ($\hat{t}^{\phantom{\dagger}}_{\nu,\alpha}$) denote triplet creation (annihilation) operators on dimer $\nu$ with magnetic quantum number $\alpha\in\{-1,0+1\}$, i.e.~$|t_\alpha\rangle\equiv \hat{t}^{\dagger}_{\alpha} |s\rangle$ on a single dimer. The latter operators are used to split the intra-dimer interactions, proportional to $x$, into operators $\hat{T}_{n}$ so that $\hat{T}_{n}$ contains all processes, which change the number of triplets by $n\in\{-1,0,1\}$. For the  Ising-Heisenberg model, all operators with $n=0$ represent triplet-triplet interactions while operators with $n=1$ ($n=-1$) create (destroy) a triplet if a second triplet is present on an appropriate nearest-neighbor dimer. 

The essential goal is now to transform Eq.~(\ref{eqn:Hpcut}) into an effective model conserving the number of triplons so that the effective Hamiltonian after the continuous unitary transformation commutes with the counting operator $\hat{Q}$. Triplons with total spin one are the elementary excitations of coupled-dimer systems and can be viewed as triplets dressed with a polarization cloud.\cite{schm03} In a finite magnetic field, the relevant processes for the magnetization process above the singlet-dimer phase have maximum values of total $S^z$ as long as bound states of triplons with different quantum numbers do not become relevant at low energies.\cite{manm11} Here we focus on this channel, but we stress that also all other channels with different quantum numbers could be calculated within the same framework. The general form of the effective low-energy model is then given by
\begin{eqnarray}
\frac{\mathcal{H}_{\rm eff}}{J} &=& E_0+\frac{h}{J} \hat{H}_{h} +\sum_{i,\delta} t_\delta ^{o}\, b^\dagger_{i+\delta} b^{\phantom{\dagger}}_{i} 
\nonumber\\
&+& \sum_{i,\delta_n} V_{\delta_1,\delta_2,\delta_3}^{o} \, b^\dagger_{i+\delta_3} b^\dagger_{i+\delta_2} b^{\phantom{\dagger}}_{i+\delta_1}b^{\phantom{\dagger}}_{i} \,\ldots\, ,
\label{eq:ham_eff}
\end{eqnarray}
where the sums run over the sites $i$ of the effective square lattice built by dimers of the Shastry-Sutherland model and $o\in\{{\rm v},{\rm h}\}$ gives the orientation {\it vertical} or {\it horizontal} of dimer $i$. The dots "$\ldots$" represent terms containing more than four operators. The hardcore boson operator $b^\dagger_i$ ($b^{\phantom{\dagger}}_{i}$) corresponds to the creation (annihilation) of a triplet $|t^1\rangle$ on dimer $i$.  Note that the constant $E_0$ is the same as the one before the continuous unitary transformation since the product state of singlets is an exact eigenstate. 

Often the effective low-energy model is derived as a high-order series expansion in $x=J'/J$ using the pCUT method. For the problem at hand, this is only possible for $\Delta=1$ where the unperturbed part of the Hamiltonian ($x=0$) has an equidistant spectrum. To this end the amplitudes of the effective model are determined in the thermodynamic limit by exploiting the linked-cluster theorem, i.e.~calculations on finite clusters are sufficient in order to treat all quantum fluctuations of a finite perturbative order correctly. The conventional quantum Heisenberg model on the Shastry-Sutherland is already special with regards to the linked-cluster expansion since the singlet ground state is an exact eigenstate and therefore no quantum fluctuations are present in the ground state. This is different for excitations, since triplets can be excited on neighboring dimers if a triplet is already present. As a consequence, there exist virtual fluctuations of triplets whose spatial extension scales with the perturbative order. The resulting effective hardcore boson model therefore contains quantum fluctuations, e.g.~correlated hopping terms or many-body interactions, to arbitrary distances and the problem cannot be solved exactly \cite{dori08,folt14}.

This is fundamentally different for the Ising-Heisenberg model. Here one has exact local conservation laws since the magnetic quantum number on dimers is a conserved quantity. This has dramatic consequences: First, single triplets $|t_1\rangle$ remain static. The only quantum fluctuation existing is the conversion of singlets into triplets $|t_0\rangle$ for two of the four dimers being nearest neigbors of a triplet $|t_1\rangle$. Quantum fluctuations are therefore confined to nearest-neighbor dimers and the extension does not scale with the perturbative order. Second, the operator $\hat{T}_0$ does not link different dimers. As a consequence, the range and the number of operators in the effective model Eq.~(\ref{eq:ham_eff}) is finite which sets the basis for an exact solution.

All amplitudes of the effective model can be determined on graphs consisting of at most three neighboring dimers in $x$- or $y$-direction, i.e.~the number and the size of graphs is tiny. One can therefore derive these amplitudes easily by pCUTs as a series expansion in $J'/J$ for $\Delta=1$. More importantly, one can determine the contibutions exactly on the finite set of graphs with gCUTs for any value of $\Delta$ which we would like to examplify for the chemical potential $\mu$. First, there is the local contribution $\mu^{(1)}\equiv\langle t_1|\hat{Q}|t_1\rangle=(1+\Delta)/2$ to the chemical potential. The only other contribution to the chemical potential can be calculated on a graph of two nearest-neighbor dimers, e.g.~oriented in $x$-direction such that the left dimer is a horizonzal one. As a bare one-particle reference state we take $|0\rangle\equiv|s\rangle|t_1\rangle$. The intra-dimer interaction $\sum_n\hat{T}_n$ creates only the single state $|1\rangle\equiv|t_0\rangle|t_1\rangle$ so that the calculation reduces to the diagonalization of the single 2x2 matrix $\langle\alpha^\prime|\langle\beta^\prime |\hat{H}|\alpha\rangle|\beta\rangle$ with $\alpha^{(\prime)},\beta^{(\prime)}\in\{0,1\}$. The lowest eigenvalue of this matrix is $\Delta+\frac{1}{2}-\frac{1}{2}\sqrt{\Delta^2+x^2}$ which corresponds to the sum $\mu^{(1)}+\mu^{(2)}$. Exactly the same kind of contribution is obtained for the two-dimer graph in $x$-direction such that the triplet $|t_1\rangle $ is located on a vertical dimer which is left. Therefore, the complete expression for the chemical potential in the thermodynamical limit is given non-perturbatively by
\begin{equation}
 \mu\equiv \mu^{(1)}+2\mu^{(2)} = \frac{1+3\Delta}{2}-\sqrt{\Delta^2+x^2}\quad ,
\end{equation}
which reduces to the pCUT expresion for $\Delta=1$ when performing a Taylor series in $x$ (see below). The same kind of reasoning can be done for all other contributions to the effective model. One obtains
\begin{widetext}
\begin{equation}
\label{eqn:H05}
\frac{{\hat H}_{\rm eff}}{J} = \mu\sum_i \hat{n}_i +V_1\sum_{\langle i,j\rangle} \hat{n}_i\hat{n}_j +V_3\left[\sum_{i\,{\rm vertical}} \hat{n}_i(1-\hat{n}_{i+e_x})\hat{n}_{i+2e_x}+ \sum_{i\, {\rm horizontal}} \hat{n}_i(1-\hat{n}_{i+e_y})\hat{n}_{i+2e_y} \right]\quad ,
\end{equation}
\end{widetext}
where all three amplitudes are given exactly by
\begin{eqnarray}
 \mu &=&  \frac{1+3\Delta}{2}-\sqrt{\Delta^2+x^2} \nonumber\\
     &=& \frac{1+\Delta}{2}-\frac{1}{2\Delta}x^2 +\frac{1}{8\Delta^3} x^4 \quad \ldots
\end{eqnarray}
\begin{eqnarray}
 V_1 &=& -\frac{\Delta-x}{2}+\frac{1}{2}\sqrt{\Delta+x^2} \nonumber\\
      &=& \frac{1}{2}x + \frac{1}{4\Delta}x^2-\frac{1}{16\Delta^3}x^4\quad \ldots
\end{eqnarray}
\begin{eqnarray}
 V_3 &=& -\Delta+\sqrt{\Delta^2+x^2}  \nonumber\\ 
      &=& \frac{1}{2\Delta}x^2 - \frac{1}{8\Delta^3} x^4 \quad \ldots \quad .
\end{eqnarray}
The effective model is purely classical since it solely consists of diagonal operators such as the chemical potential and density-density interactions. Besides the two repulsive two-particle interactions, there is also one attractive three-body interaction for three neighboring particles in $x$- or $y$-direction depending on the orientation. The leading perturbative order of all couplings corresponds to the ones of the full quantum Heisenberg model on the Shastry-Sutherland lattice for $\Delta=1$ \cite{dori08}. The essential difference is the absence of off-diagonal operators such as correlated hopping processes which introduce quantum fluctuations in the effective low-energy model. Finally, it might be convenient to relate the interactions to the chemical potential which gives $V_1=\frac{1}{2}(1+x-\mu)$ and $V_3=1-\mu$.

\subsection{Correspondence}

Both formulations are equivalent if we notice the correspondence between spin states in (\ref{h-2d-od}) and hard-core bosons in (\ref{eqn:H05}). The empty site $(i,j)$ in the particle formulation corresponds to the spin configuration \mbox{$s_{1,i,j}^z=\frac{1}{2}$}, $s_{2,i,j}^z=-\frac{1}{2}$ on a dimer, while occupied sites have \mbox{$s_{1,i,j}^z=\frac{1}{2}$}, $s_{2,i,j}^z=\frac{1}{2}$. 

Let us briefly describe the correspondence between eigenstates of the initial and diagonalized local cluster Hamiltonians, the latter being diagonal in the basis spanned over four eigenstates of the spin operators $s_{1,i}^z$ and $s_{2,i}^z$. Applying the inverse unitary transformation one obtains the following relations between the relevant eigenstates:
\begin{eqnarray}
\label{U_states}
&&|\tilde\uparrow_{1,i,j}\tilde\uparrow_{2,i,j}\rangle
  =U_{i,j}^{+}|\uparrow_{1,i,j}\uparrow_{2,i,j}\rangle=|\uparrow_{1,i,j}\uparrow_{2,i,j}\rangle, 
\nonumber\\
&&|\tilde\downarrow_{1,i,j}\tilde\downarrow_{2,i,j}\rangle
=U_{i,j}^{+}|\downarrow_{1,i,j}\downarrow_{2,i,j}\rangle=|\downarrow_{1,i,j}\downarrow_{2,i,j}\rangle, 
\nonumber\\
&&|\tilde\uparrow_{1,i,j}\tilde\downarrow_{2,i,j}\rangle
=U_{i,j}^{+}|\uparrow_{1,i,j}\downarrow_{2,i,j}\rangle
\nonumber\\
&&=\cos\frac{\alpha_{i,j}}{2}|\uparrow_{1,i,j}\downarrow_{2,i,j}\rangle
-\sin\frac{\alpha_{i,j}}{2}|\downarrow_{1,i,j}\uparrow_{2,i,j}\rangle,
\nonumber\\
&&|\tilde\downarrow_{1,i,j}\tilde\uparrow_{2,i,j}\rangle
=U_{i,j}^{+}|\downarrow_{1,i,j}\uparrow_{2,i,j}\rangle
\nonumber\\
&&=\sin\frac{\alpha_{i,j}}{2}|\uparrow_{1,i,j}\downarrow_{2,i,j}\rangle
+\cos\frac{\alpha_{i,j}}{2}|\downarrow_{1,i,j}\uparrow_{2,i,j}\rangle. 
\end{eqnarray}
Note that the mixing angle $\alpha_{i,j}$ entering the two antiferromagnetic eigenstates of the central dimer depends, according to Eq.~(\ref{tp}), just on the difference between the total spin of neighboring dimers. The first two polarized triplet states with total spin $S_{i,j}^z=\pm1$ are not affected at all by the unitary transformation since the initial local cluster Hamiltonian was diagonal in this particular subspace, while the other two antiferromagnetic states with total spin $S_{i,j}^z=0$ are quantum-mechanically mixed by the unitary transformation. As a result, the classical antiferromagnetic states $|\tilde\uparrow_{1,i,j}\tilde\downarrow_{2,i,j}\rangle$ and $|\tilde\downarrow_{1,i,j}\tilde\uparrow_{2,i,j}\rangle$ of the diagonalized cluster Hamiltonian correspond to the quantum antiferromagnetic order that is subject to a quantum reduction of the magnetization given by $\langle s_{1,i,j}^z\rangle=-\langle s_{2,i,j}^z\rangle=\pm\frac{1}{2}\cos\alpha_{i,j}$
($\langle s_{1,i,j}^z\rangle=-\langle s_{2,i,j}^z\rangle=\mp\frac{1}{2}\cos\alpha_{i,j}$) in the state $|\tilde\uparrow_{1,i,j}\tilde\downarrow_{2,i,j}\rangle$ ($|\tilde\downarrow_{1,i,j}\tilde\uparrow_{2,i,j}\rangle$).

\section{Results and discussion}
\label{sec-gs}
In this section, we report all exact ground states of the spin-1/2 Ising-Heisenberg model on the Shastry-Sutherland lattice, which will be subsequently used for constructing the complete ground-state phase diagram in a field. In what follows, we consider only the particular case of antiferromagnetic interactions $J>0$ and $J'\geq 0$. The $z$-component of the $XXZ$ Heisenberg intra-dimer coupling will be used as the energy unit by setting $J=1$. 

First, it can be easily checked that the diagonal form of the local cluster Hamiltonian (\ref{ham-local1}) has always lower energy for the spin configuration $s_{1,i,j}^z=\frac{1}{2}$, $s_{2,i,j}^z=-\frac{1}{2}$ than for the reverse spin configuration $s_{1,i,j}^z=-\frac{1}{2}$, $s_{2,i,j}^z=\frac{1}{2}$ provided the exchange anisotropy $\Delta > 0$. Hence, it follows that the latter antiferromagnetic state can be thoroughly excluded from further considerations when looking for the lowest-energy eigenstates of the spin-1/2 Ising-Heisenberg model on the Shastry-Sutherland lattice, but it cannot be neglected in the special Ising limit $\Delta=0$. To find all possible ground states of the Ising-Heisenberg model with $\Delta>0$, it is therefore sufficient to consider all spin configurations accessible entirely from three states of the diagonalized local cluster Hamiltonians: $s_{1,i,j}^z=s_{2,i,j}^z=\frac{1}{2}$, $s_{1,i,j}^z=-s_{2,i,j}^z=\frac{1}{2}$, and $s_{1,i,j}^z=s_{2,i,j}^z=-\frac{1}{2}$, which can be alternatively identified as fictitious spin states $S_{i,j}^z=+1,0,-1$ of some classical effective spin-1 model. The diagonal form of the Hamiltonian (\ref{h-2d-od})-(\ref{h-2d-odd}) can be then rewritten into the following form:
\begin{eqnarray}
&&\tilde{H}=\sum_{i,j=1}^{N} \tilde{H}_{i,j}^{0}
+\sideset{}{'}\sum_{i,j=1}^N \tilde{V}_{i,[j-1:j+1]}
+ \sideset{}{''}\sum_{i,j=1}^N \tilde{V}_{[i-1:i+1],j},
\nonumber \\
&&\tilde{H}_{i,j}^0=
-\frac{1+\Delta}{2}J[1-(S_{i,j}^z)^2] +\frac{J}{4} - hS_{i,j}^z,
\nonumber\\
&&\tilde{V}_{[i-1:i+1],j}{=}
-\frac{1}{2}[1-(S_{i,j}^z)^2]I(|S_{i+1,j}^z-S_{i-1,j}^z|)
\nonumber\\
&&\qquad\qquad\quad+\frac{J'}{2}S_{i,j}^z(S_{i+1,j}^z+S_{i-1,j}^z),
\nonumber\\
&&\tilde{V}_{i,[j-1:j+1]}{=}
-\frac{1}{2}[1-(S_{i,j}^z)^2]I(|S_{i,j+1}^z-S_{i,j-1}^z|)
\nonumber\\
&&\qquad\qquad\quad+\frac{J'}{2}S_{i,j}^z(S_{i,j+1}^z+S_{i,j-1}^z).
\label{ham-eff}
\end{eqnarray}
It should be noted that the analogous spin-1 representation is also valid for the particular case with $\Delta=0$, which corresponds to the purely classical spin-1/2 Ising model on the Shastry-Sutherland lattice. Unlike the previous case, the effective staggered field $\Delta J/2$ completely vanishes in the limiting Ising case $\Delta=0$ and consequently, two antiferromagnetic 
states $s_{1,i,j}^z=-s_{2,i,j}^z=\pm\frac{1}{2}$ can have equal energies unless the effective interaction among three consecutive dimers makes the energy of the antiferromagnetic state $s_{1,i,j}^z=-s_{2,i,j}^z=\frac{1}{2}$ lower. Thus, the two-fold degeneracy of the antiferromagnetic states $s_{1,i,j}^z=-s_{2,i,j}^z=\pm\frac{1}{2}$ on all dimers can lead to a highly degenerate ground-state manifold for the spin-1/2 Ising model on the Shastry-Sutherland lattice in contrast to the spin-1/2 Ising-Heisenberg model with $\Delta>0$.

By inspection, we have found by minimizing the effective Hamiltonian (\ref{ham-eff}) six distinct ground states (see Fig.~\ref{fig_phases} for a schematic illustration of individual ground states):
\begin{figure*}[p]
\epsfig{file=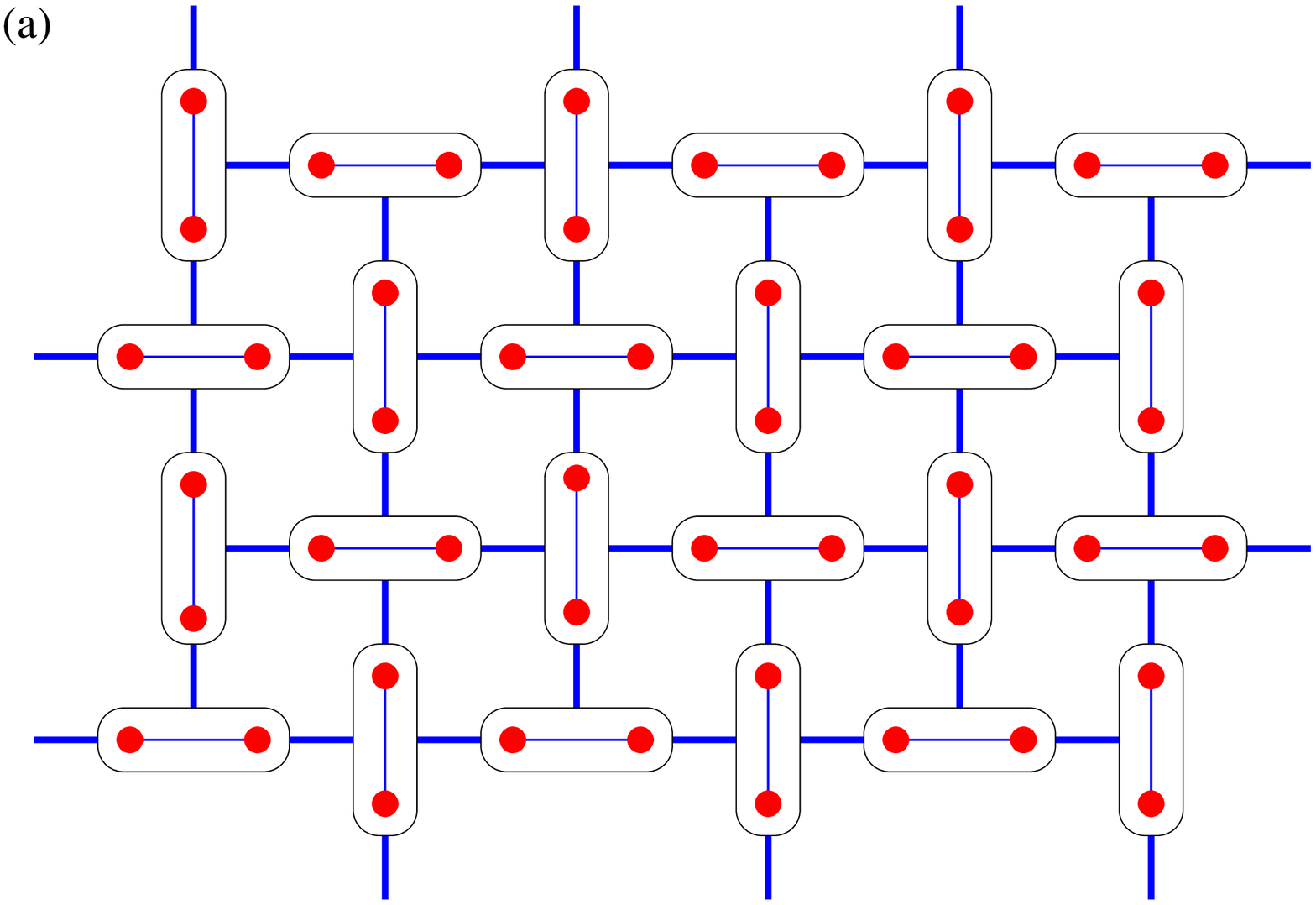, width=0.85\columnwidth}
\epsfig{file=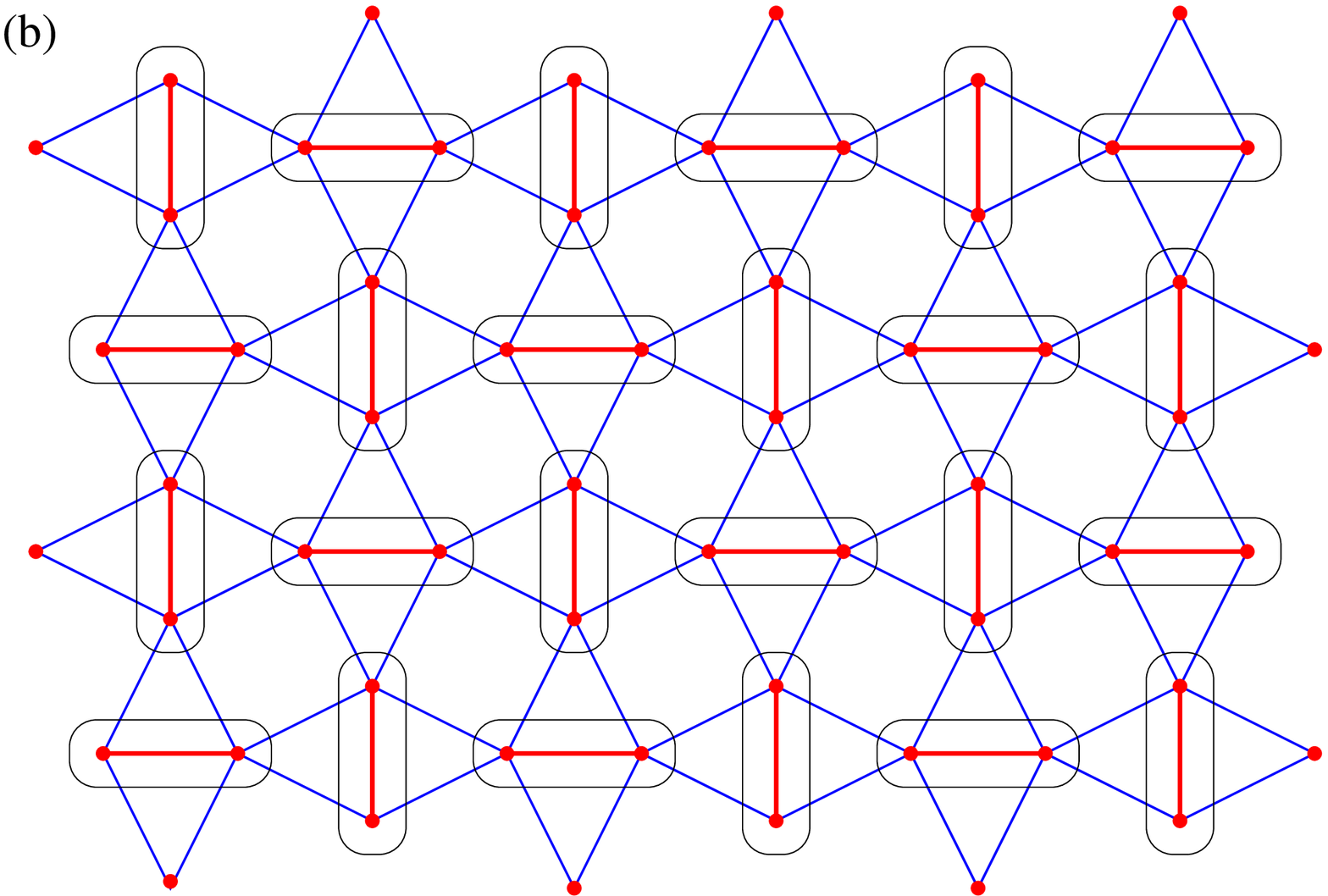, width=0.85\columnwidth}\\
\epsfig{file=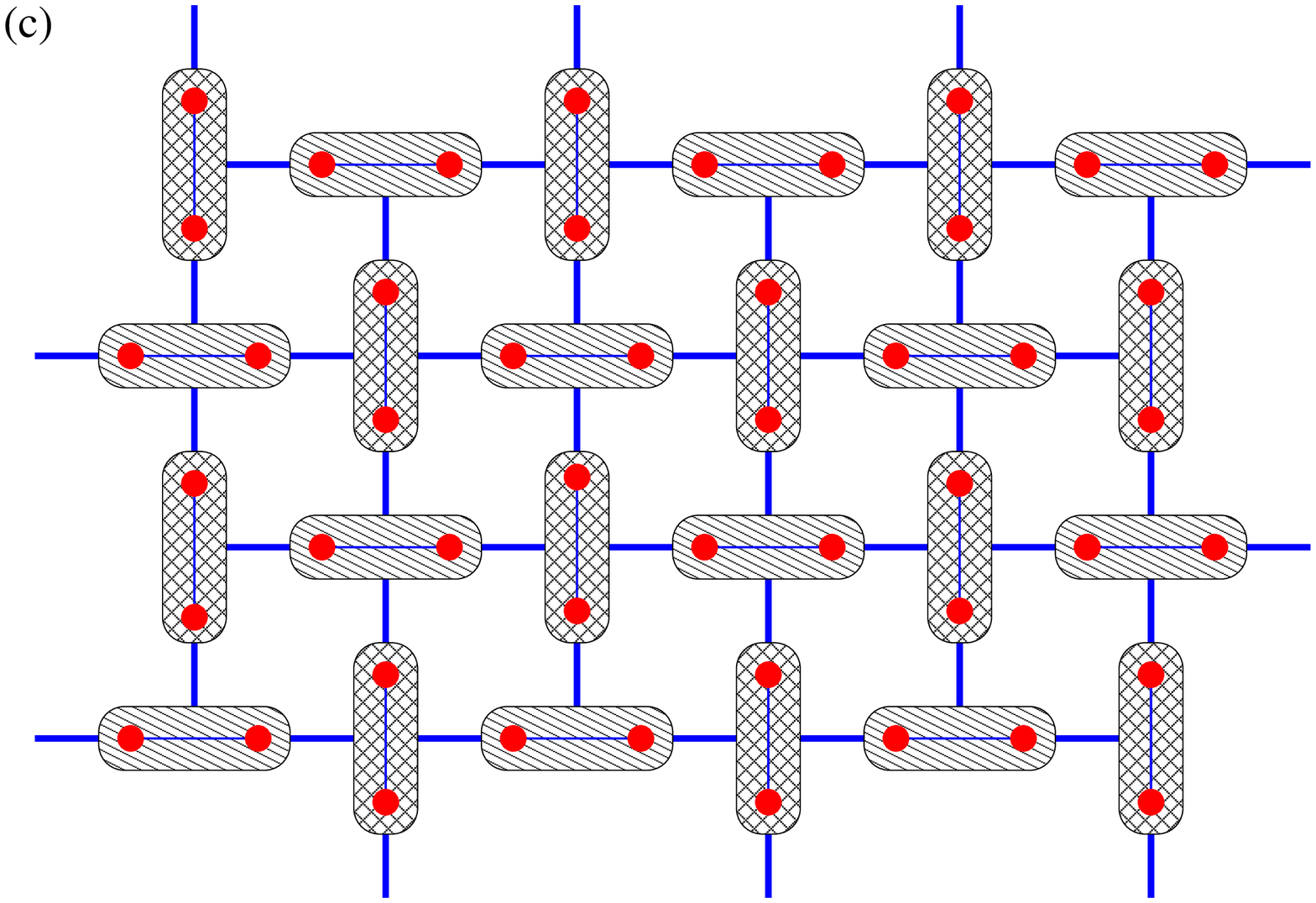, width=0.85\columnwidth}
\epsfig{file=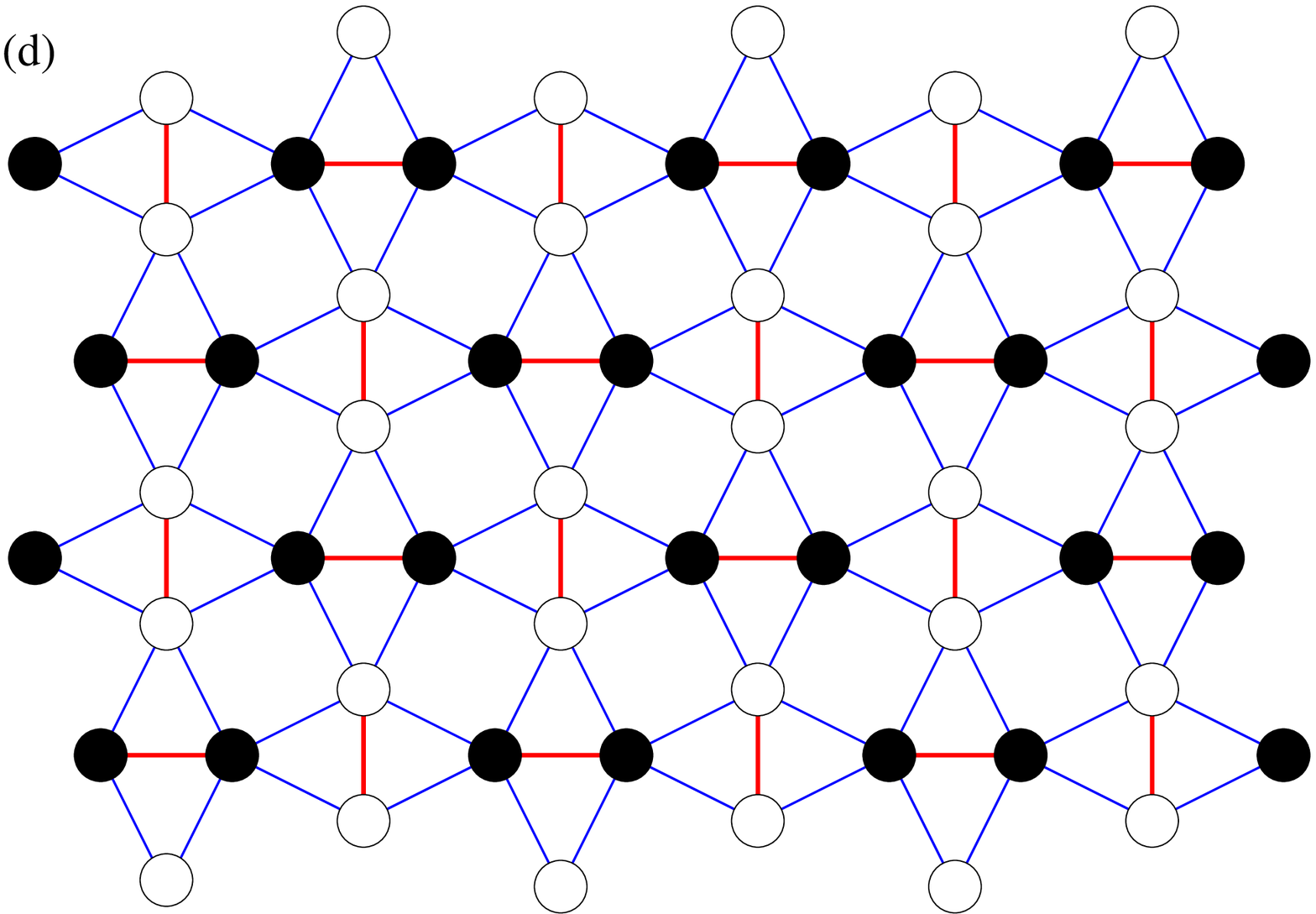, width=0.85\columnwidth}\\
\epsfig{file=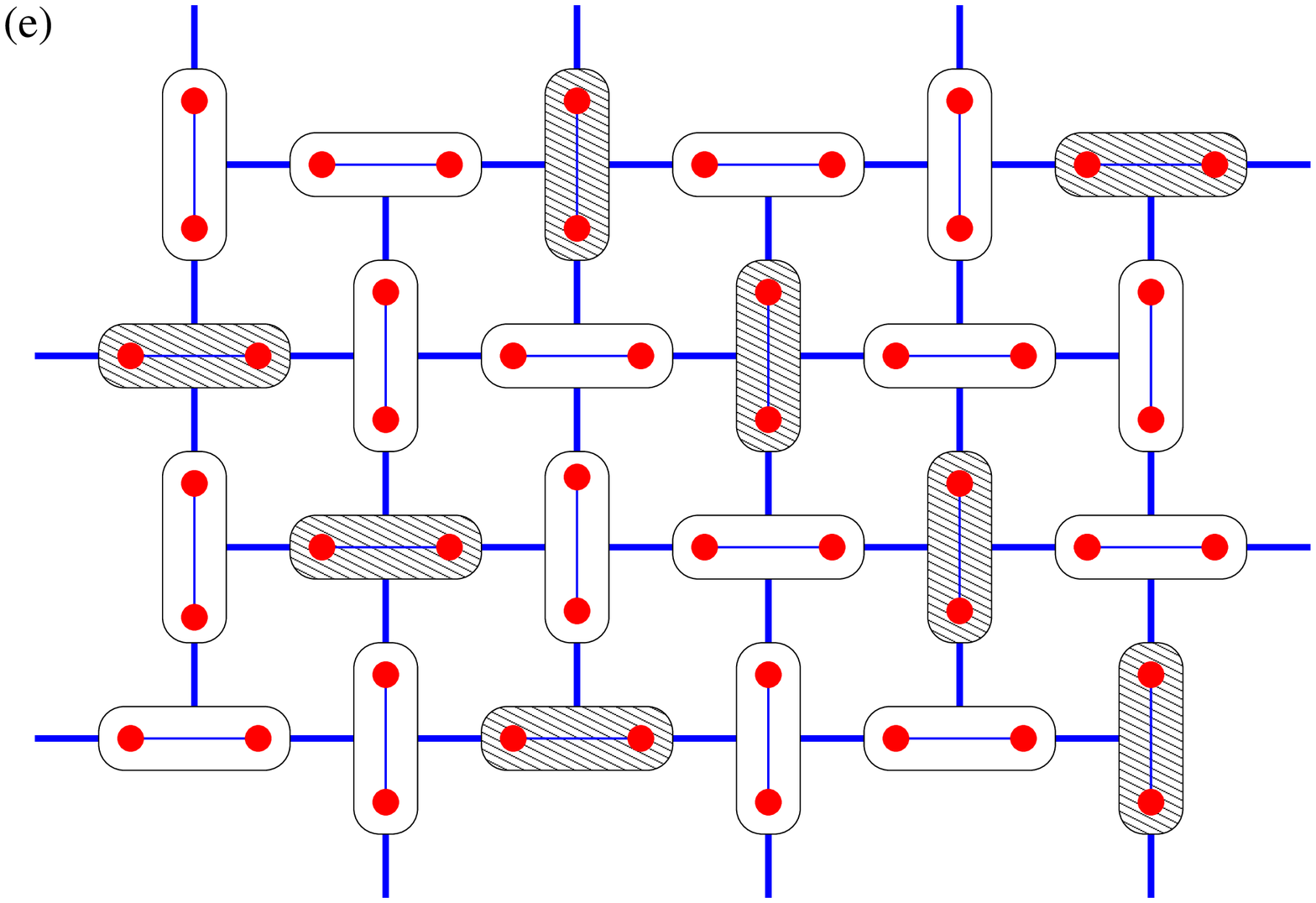, width=0.85\columnwidth}
\epsfig{file=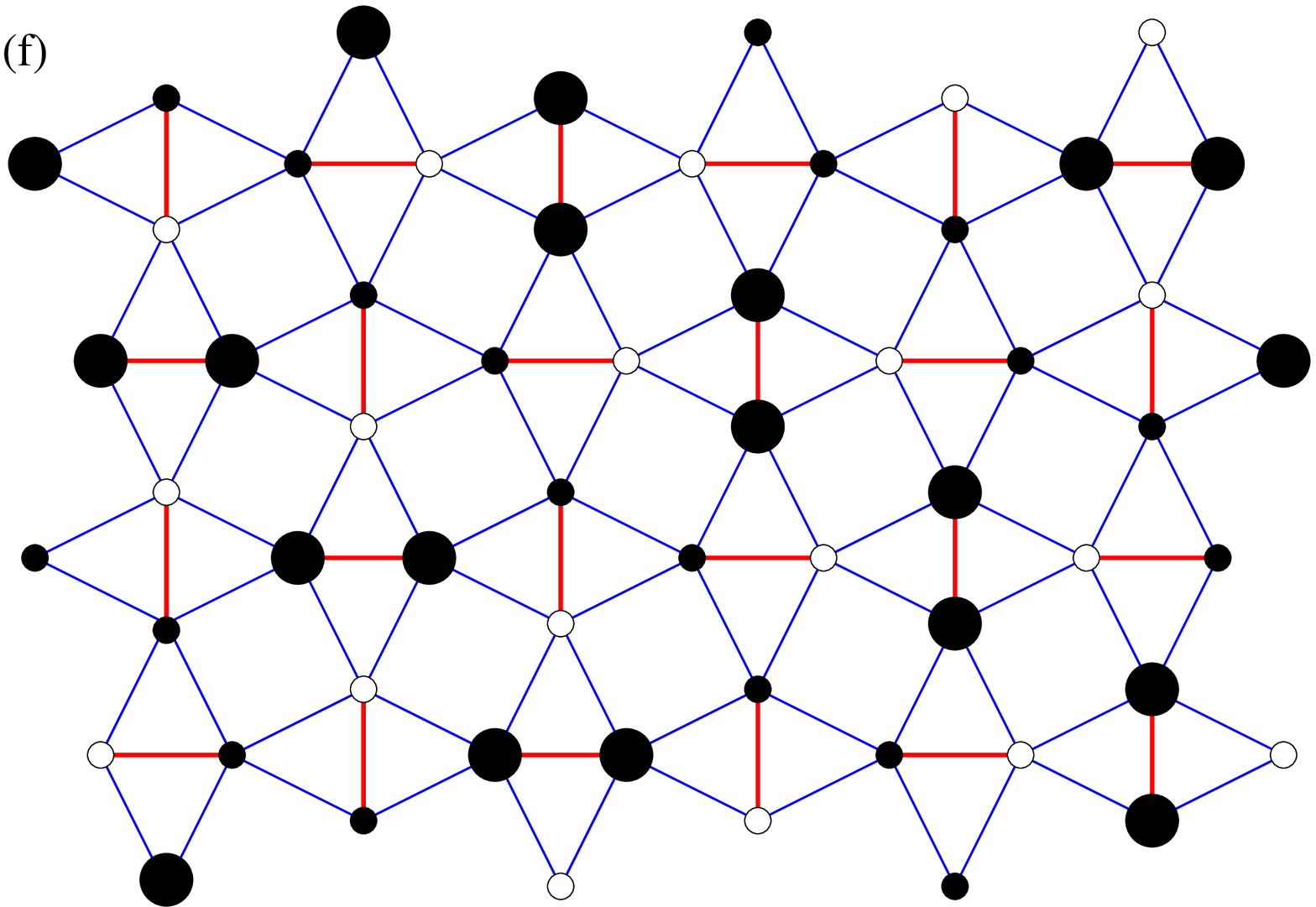, width=0.85\columnwidth}\\
\epsfig{file=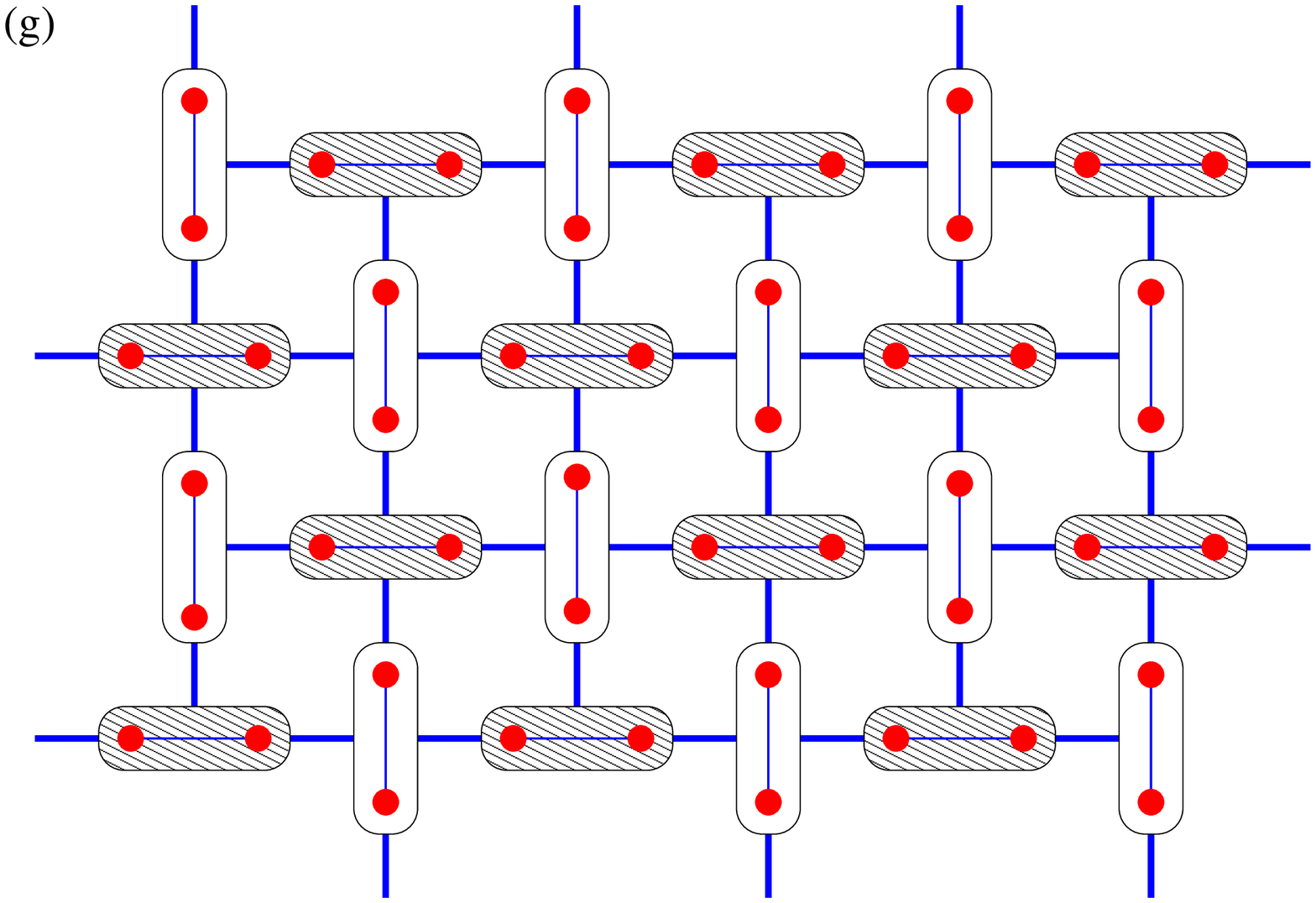, width=0.85\columnwidth}
\epsfig{file=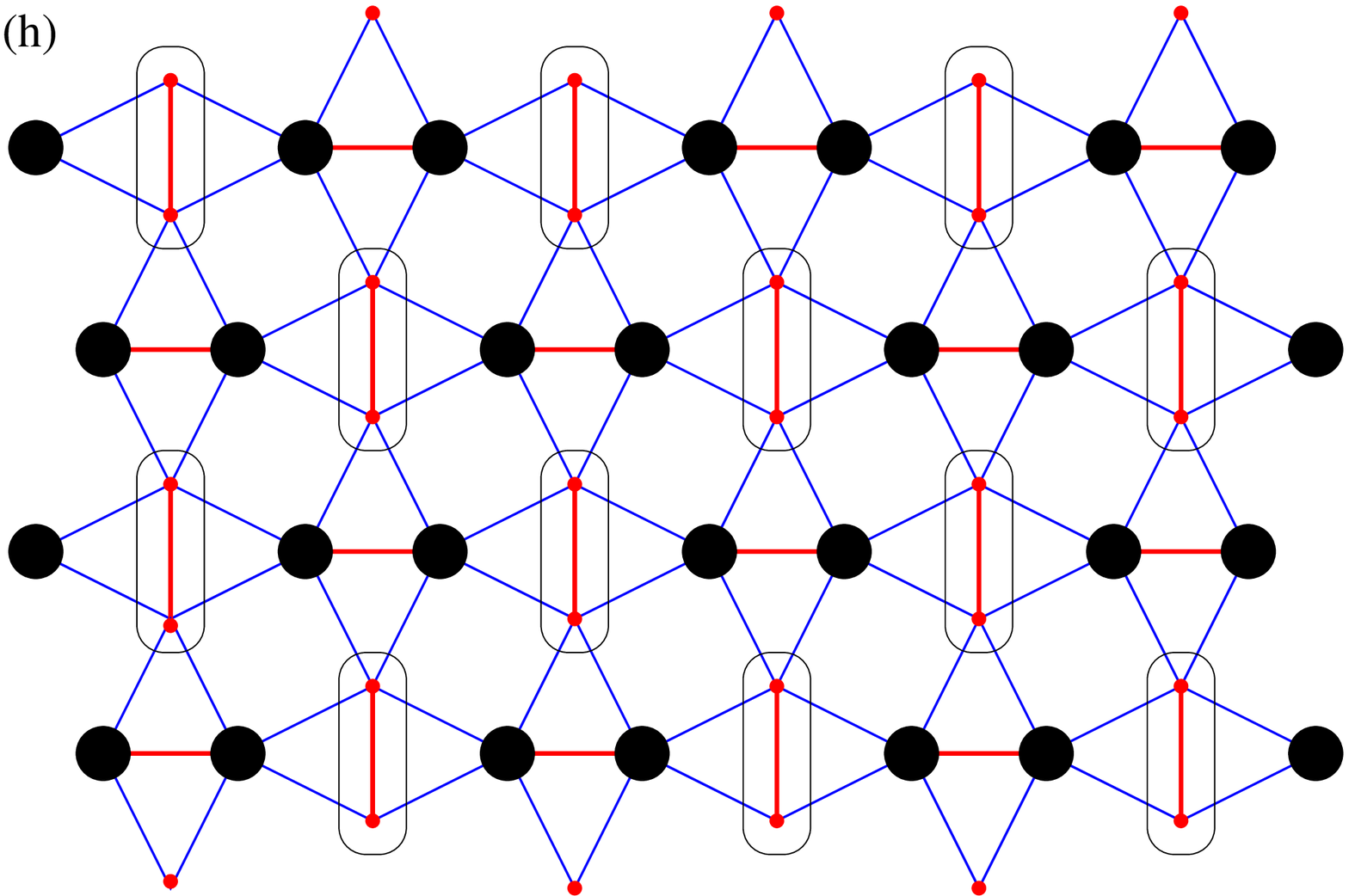, width=0.85\columnwidth}
\caption{(Color online) Schematic representation of exact ground states of the spin-1/2 Ising-Heisenberg model on the Shastry-Sutherland lattice. The left and right panels show spin arrangements relevant to the  classical effective spin model given by the Hamiltonian (\ref{ham-eff}) and the original quantum spin model defined through the Hamiltonian (\ref{ham0}), respectively. The rows from top to bottom correspond to the singlet-dimer phase, the antiferromagnetic phase, the stripe 1/3-plateau phase and the checkerboard 1/2-plateau phase. Shaded (transparent) dimers on the left panel denote the polarized triplet (singlet) states. On the right panel, the ellipse denotes a singlet-dimer state, filled circles denote spins oriented in a direction of the external magnetic field, empty circles denote spins oriented in opposite to the external magnetic field, whereas the reduced diameter of both kinds of circles corresponds to the quantum reduction of the local magnetization (\ref{qrm}).}
\label{fig_phases}
\end{figure*}
\begin{itemize}
\item the unique singlet-dimer (SD) phase constituted by a direct product over singlet-dimer states on the Heisenberg dimers (see Fig.~\ref{fig_phases}(a),(b)):
 \begin{eqnarray}
 \label{SD-phase}
&& |\mbox{SD}\rangle=\prod_{i,j}^N |\tilde{0}_{i,j}\rangle
  =\prod_{i,j}^N |{\cal S}_{i,j}\rangle,
  \\
&& |{\cal S}_{i,j}\rangle=\frac{1}{2}(|\uparrow_{1,i,j}\downarrow_{2,i,j}\rangle - |\downarrow_{1,i,j}\uparrow_{2,i,j}\rangle).
 \end{eqnarray}
\item the highly degenerate Ising-dimer (ID) phase constituted by a direct product over two-fold degenerate antiferromagnetic states $|\uparrow_{1,i,j}\downarrow_{2,i,j}\rangle$ and $|\downarrow_{1,i,j}\uparrow_{2,i,j}\rangle$ on all dimers. This ground state has a high macroscopic degeneracy proportional to the total number of dimers $2^{N^2}$ and it only exists in the Ising limit $\Delta=0$.
\item the antiferromagnetic (AF) phase formed by a direct product over two kinds of polarized triplet states $S_{i,j}^z=\pm1$ of the dimers, which regularly alternate in a such way that each dimer is polarized in opposite direction with respect to all its nearest-neighbor dimers (see Fig.~\ref{fig_phases}(c),(d)):
 \begin{eqnarray}
 \label{Af-phase}
 |{\rm AF}\rangle
 &=&
 \left[\sideset{}{'}\prod_{i,j=1}^{N}|\tilde\downarrow_{i,j}\rangle\right] 
 \left[\sideset{}{''}\prod_{i,j=1}^{N}|\tilde\uparrow_{i,j}\rangle\right]
 \nonumber\\ 
 &=&
\left[\sideset{}{'}\prod_{i,j=1}^{N}|\downarrow_{1,i,j}\downarrow_{2,i,j}\rangle\right]
\left[\sideset{}{''}\prod_{i,j=1}^{N}|\uparrow_{1,i,j}\uparrow_{2,i,j}\rangle\right].
 \end{eqnarray}
In these expressions, the symbols $\prod'$ and $\prod''$ denote the products over all vertical dimers $(i+j)=odd$ and all horizontal dimers $(i+j)=even$, respectively. The AF ground state is doubly degenerate, because another state can be created from the eigenstate (\ref{Af-phase}) by inter-changing the states of the horizontal and vertical dimers.
\item the stripe 1/3-plateau phase in which each diagonal stripe of the polarized dimers regularly alternates with two stripes of dimers in a spin-singlet-like (non-magnetic) state 
(see Fig.~\ref{fig_phases}(e),(f)):
 \begin{eqnarray}
 \label{1o3-phase}
 && |m=1/3\rangle 
 =\left[\sideset{}{'}\prod_{i,j=1}^{N}|\tilde\uparrow_{i,j}\rangle\right]
 \left[\sideset{}{''}\prod_{i,j=1}^{N}|\tilde0_{i,j}\rangle\right]
 \left[\sideset{}{'''}\prod_{i,j=1}^{N}|\tilde0_{i,j}\rangle\right]
 \nonumber\\
 && =\left[\sideset{}{'}\prod_{i,j=1}^{N}|\uparrow_{1,i,j}\uparrow_{2,i,j}\rangle\right]
 \left[\sideset{}{''}\prod_{i,j=1}^{N}|\phi^{(-)}_{i,j}\rangle\right]
 \left[\sideset{}{'''}\prod_{i,j=1}^{N}|\phi^{(+)}_{i,j}\rangle\right],
 \nonumber\\
 \\
 && |\phi^{(\pm)}_{i,j}\rangle=\cos\frac{\alpha^{(\pm)}}{2}|\downarrow_{1,i,j}\uparrow_{2,i,j}\rangle 
 - \sin\frac{\alpha^{(\pm)}}{2}|\uparrow_{1,i,j}\downarrow_{2,i,j}\rangle. 
 \nonumber\\
  \label{singletlike}
 \end{eqnarray}
Here, the symbols $\prod'$, $\prod''$, $\prod'''$ denote products over indices $i + j=3L+1, 3L+2, 3L$ or $i - j=3L+1, 3L+2, 3L$ ($L$ is any integer), the mixing angle $\alpha^{(\pm)}$ in the spin-singlet-like states $|\phi^{(\pm)}_{i,j}\rangle$ is defined as $\alpha^{(\pm)}= \arctan (\pm \Delta J/J')$ with $\alpha \in [0, \pi]$. The spin-singlet-like states capture the quantum antiferromagnetic order on the Heisenberg dimers, which can be characterized by a non-zero but not fully saturated staggered magnetization related to the quantum reduction of the local magnetizations depending on the mutual competition between the Ising inter-dimer interaction and the transverse $XX$-part of the Heisenberg intra-dimer interaction
\begin{eqnarray}
\langle s_{1,i,j}^z\rangle_{\phi^{(\pm)}} = - \langle s_{2,i,j}^z \rangle_{\phi^{(\pm)}} = \pm \frac{1}{2} \frac{J'}{\sqrt{\Delta^2 J^2 + J'^{2}}},
\label{qrm}
\end{eqnarray}
where $\langle \cdots\rangle_{\phi^{(\pm)}}=\langle\phi^{(\pm)}_{i,j}|\cdots|\phi^{(\pm)}_{i,j}\rangle$. This ground state is six-fold degenerate, since other five states can be created from the eigenstate (\ref{1o3-phase}) by translation and/or reflection. 
\item the checkerboard 1/2-plateau phase in which the singlet-dimer state on the vertical dimers regularly alternates with the polarized state on the horizontal dimers or vice versa (see Fig.~\ref{fig_phases}(g),(h)):
  \begin{eqnarray}
 \label{1o2-phase}
 |m=1/2\rangle&=&
 \left[\sideset{}{'}\prod_{i,j=1}^{N}|\tilde0_{i,j}\rangle\right]
 \left[\sideset{}{''}\prod_{i,j=1}^{N}|\tilde\uparrow_{i,j}\rangle\right] 
 \nonumber\\
 &=&
 \left[\sideset{}{'}\prod_{i,j=1}^{N}|{\cal S}_{i,j}\rangle\right]
 \left[\sideset{}{''}\prod_{i,j=1}^{N}|\uparrow_{1,i,j}\uparrow_{2,i,j}\rangle\right]. 
 \end{eqnarray}
Here, the symbols $\prod'$ and $\prod''$ denote the products over all vertical dimers $(i+j)=odd$ and all horizontal dimers $(i+j)=even$, respectively. This ground state is doubly degenerate because of the possible inter-change of the states of vertical and horizontal dimers.
\item the saturated paramagnetic phase with the fully polarized dimers:
 \begin{eqnarray}
 |m=1\rangle=\prod_{i,j=1}^N |\tilde\uparrow_{i,j}\rangle
 =\prod_{i,j=1}^N |\uparrow_{1,i,j}\uparrow_{2,i,j}\rangle. 
 \label{spp} 
 \end{eqnarray}
\end{itemize}

At this stage, we can adapt the procedure developed in Refs. [\onlinecite{dubl12,dubl13}] for the classical spin-1/2 Ising model on the Shastry-Sutherland lattice in order to find out whether or not some of the exact eigenstates (\ref{SD-phase})--(\ref{spp}) represents true ground state in a particular parameter range. For this purpose, it is useful to rewrite at first the overall configurational energy as a sum of energies of local cluster Hamiltonians that depend solely on the $z$-component of the total spin on all dimers:
\begin{widetext}
\begin{eqnarray}
\label{ham-local2}
H'_{[i-1:i+1],j}(S_{i-1,j}^z,S_{i,j}^z,S_{i+1,j}^z)=
\tilde{V}_{[i-1:i+1],j} -\frac{h}{4}(S_{i-1,j}^z+2S_{i,j}^z+S_{i+1,j}^z)  - \frac{(1+2\Delta)J}{4}
\nonumber\\
+\frac{(1+\Delta) J}{2} \left\{\gamma\left[(S_{i-1,j}^z)^2+(S_{i+1,j}^z)^2\right] + (1-2\gamma)(S_{i,j}^z)^2\right\}, 
\nonumber\\
H'_{i,[j-1:j+1]}(S_{i,j-1}^z,S_{i,j}^z,S_{i,j+1}^z)=
\tilde{V}_{i,[j-1:j+1]} -\frac{h}{4}(S_{i,j-1}^z+2S_{i,j}^z+S_{i,j+1}^z)  - \frac{(1+2\Delta)J}{4}
\nonumber\\
+\frac{(1+\Delta) J}{2}\left\{\gamma\left[(S_{i,j-1}^z)^2+(S_{i,j+1}^z)^2\right] + (1-2\gamma)(S_{i,j}^z)^2\right\}, 
\end{eqnarray}
\end{widetext}
which involve a new free parameter $\gamma$ to be determined later on. Let us also introduce the simplified notation $E(S_1^z,S_2^z,S_3^z)$ for the configurational energy of the local cluster Hamiltonian (\ref{ham-local2}) involving three consecutive dimers either in a horizontal or vertical direction. 
Thus, the total energy of the model is a sum of the configurational energies $E(S_1^z,S_2^z,S_3^z)$ of all clusters (\ref{ham-local2}).
To get the ground state we have to find the cluster configurations $(S_1^z,S_2^z,S_3^z)$ which attain the lowest energy and to assemble a state of the whole system from them. All states created by such a way correspond to the ground state of the effective classical spin model. Finally, the inverse transformations $U_{i,j}^+$ (see Eq.(\ref{U_states})) is applied to recover the ground state of the initial Ising-Heisenberg model (\ref{ham-def}). 

In an absence of the external magnetic field $h=0$, we can choose the parameter $\gamma=1/2$ in order to satisfy the condition that the configurational energy $E(0,0,0)$ is the lowest for $J'<(1+\Delta)J/2$, otherwise the configurational energies $E(1,-1,1)=E(-1,1,-1)$ achieve the lowest value. In this respect, the singlet-dimer phase constitutes the zero-field ground state for $J'<(1+\Delta)J/2$, since it totally consists of the lowest-energy clusters with the configuration $(0,0,0)$. On the contrary, the lowest-energy clusters with the configurations $(1,-1,1)$ and $(-1,1,-1)$ can regularly alternate in order to produce the other zero-field antiferromagnetic ground state if the reverse condition $J'>(1+\Delta)J/2$ is met.

The situation becomes a bit more involved in non-zero magnetic field. It is clear from Fig.~\ref{fig_phases}(e),(f) that the stripe $1/3$-plateau phase can be established from the clusters (0,1,0) and (1,0,0), each of which contains just one polarized dimer from three consecutive dimers either at a central or a side position. Thus, it is necessary to verify that the energies of such clusters may become equal to each other and that they are simultaneously lowest in a certain parameter region in order to check whether or not the stripe $1/3$-plateau phase may become the ground state. The appropriate value of the parameter $\gamma$ can be therefore found according to the condition $E(0,1,0)=E(1,0,0)$:
\begin{eqnarray}
\label{gamma}
\gamma=\frac{2}{3(1+\Delta)J}\left(-\frac{h}{4}+\frac{J}{2} +\frac{1}{2}\sqrt{\Delta^2J^2+J'^2}\right).
\end{eqnarray}
The calculation of this specific value (\ref{gamma})
provides evidence that the configurational energies satisfy the inequality $E(0,0,0)<E(0,1,0)=E(1,0,0)$ for \mbox{$J'<(1+\Delta)J/2$} if the magnetic field is smaller than the first critical value ($h<h_1$)
\begin{eqnarray}
\label{h_SD-stripe}
h_1=\frac{(1+3\Delta)J}{2}-\sqrt{\Delta^2J^2+J'^2},
\end{eqnarray}
while the configurational energies obey the inequality $E(1,-1,1)=E(-1,1,-1)<E(0,1,0)=E(1,0,0)$ for $J'>J(1+\Delta)/2$ if the magnetic field is below the second critical value ($h<h_2$)
\begin{eqnarray}
\label{h_AF-stripe}
h_2=3J'-J-\sqrt{\Delta^2J^2+J'^2}.
\end{eqnarray}
It should be noted that the energies of all other configurations not mentioned in the inequalities above and below have even higher values for the considered fields and interactions, and, therefore, they are irrelevant.
These results suggest that the stripe $1/3$-plateau phase is energetically favored with respect to the singlet-dimer and antiferromagnetic phases for the magnetic fields \mbox{$h>h_1$} if $J'<(1+\Delta)J/2$ and $h>h_2$ if $J'>(1+\Delta)J/2$, respectively. Furthermore, the configurational energies pertinent to the stripe 1/3-plateau phase fulfill the condition $E(0,1,0)=E(1,0,0)<E(1,0,1)$ as long as the magnetic field does not exceed the third critical value ($h<h_3$)
\begin{eqnarray}
\label{stripe-checkerboard}
h_3=\frac{(1-3\Delta)J}{2}+2\sqrt{\Delta^2J^2+J'^2},
\end{eqnarray}
above which the checkerboard $1/2$-plateau phase develops in the ground state on account of a regular alternation of the clusters (1,0,1) and (0,1,0). 
The special value of the parameter $\gamma=1/4$ can be used to obtain the lower and upper boundaries for the checkerboard $1/2$-plateau phase defined by the conditions $E(1,0,0)<E(0,1,0)=E(1,0,1)<E(1,1,1)$. The left inequality is valid for $h>h_3$, while the right one is satisfied if the magnetic field is smaller than a fourth critical field given by
\begin{eqnarray}
\label{checkboard-polarized}
h_4=2J'+\frac{(1+\Delta)J}{2}.
\end{eqnarray}
It is quite evident from the previous argumentation that the special value of the magnetic field $h_4$ corresponds to the saturation field, above which the investigated system passes to the saturated paramagnetic phase with the fully polarized dimers into the direction of the external magnetic field.

\begin{figure*}[bt]
\epsfig{file=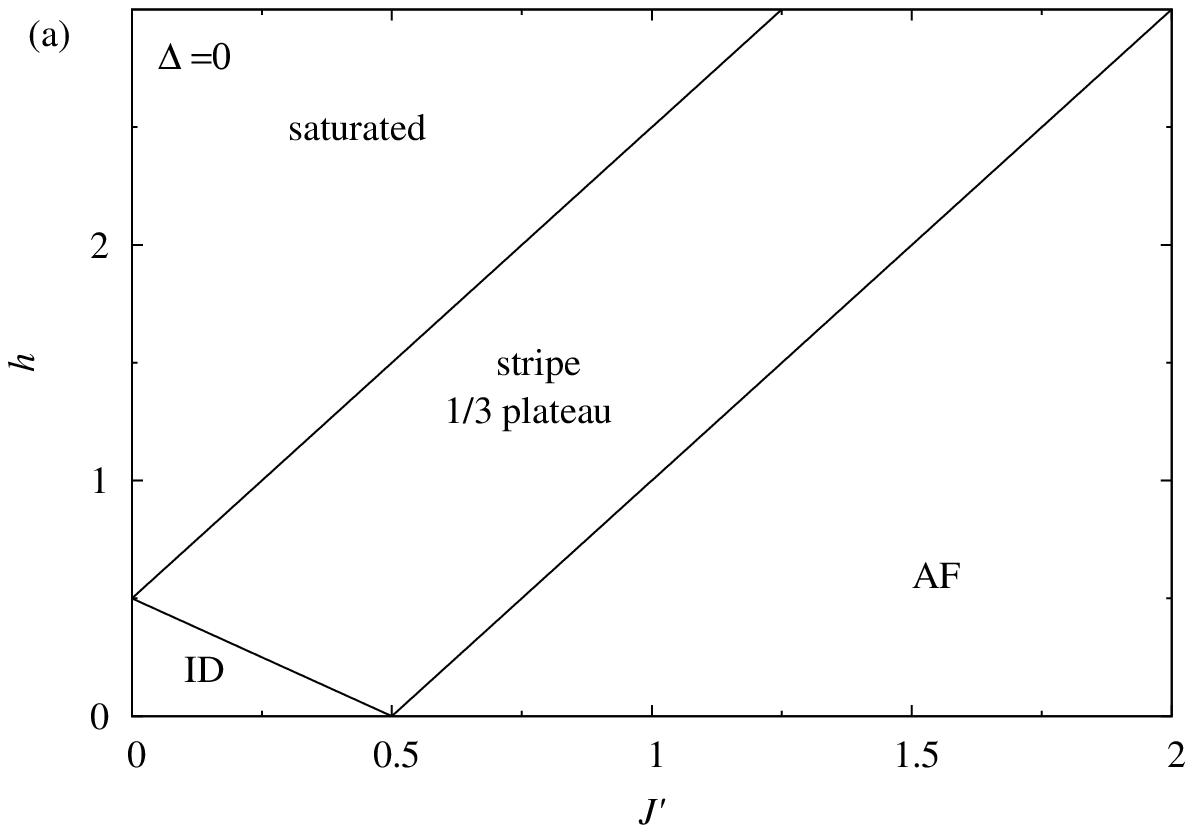, width=0.95\columnwidth}
\epsfig{file=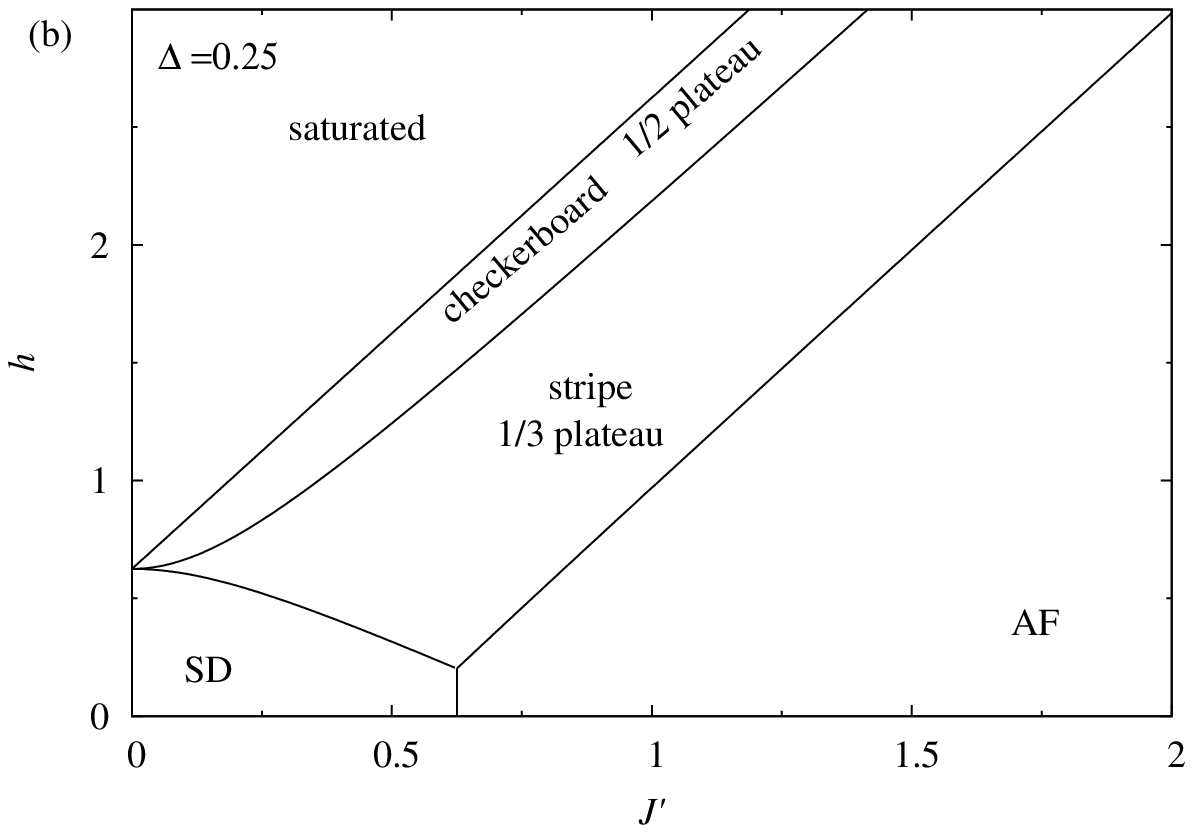, width=0.95\columnwidth}
\epsfig{file=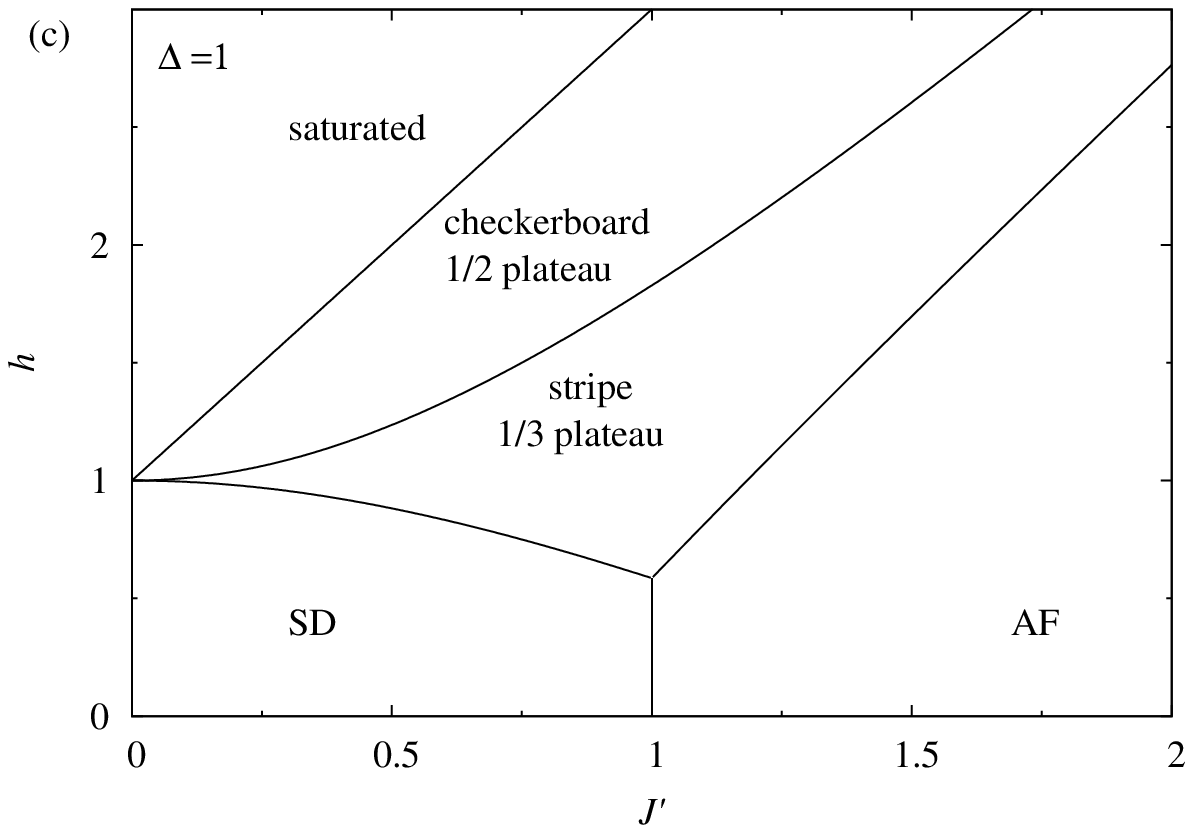, width=0.95\columnwidth}
\epsfig{file=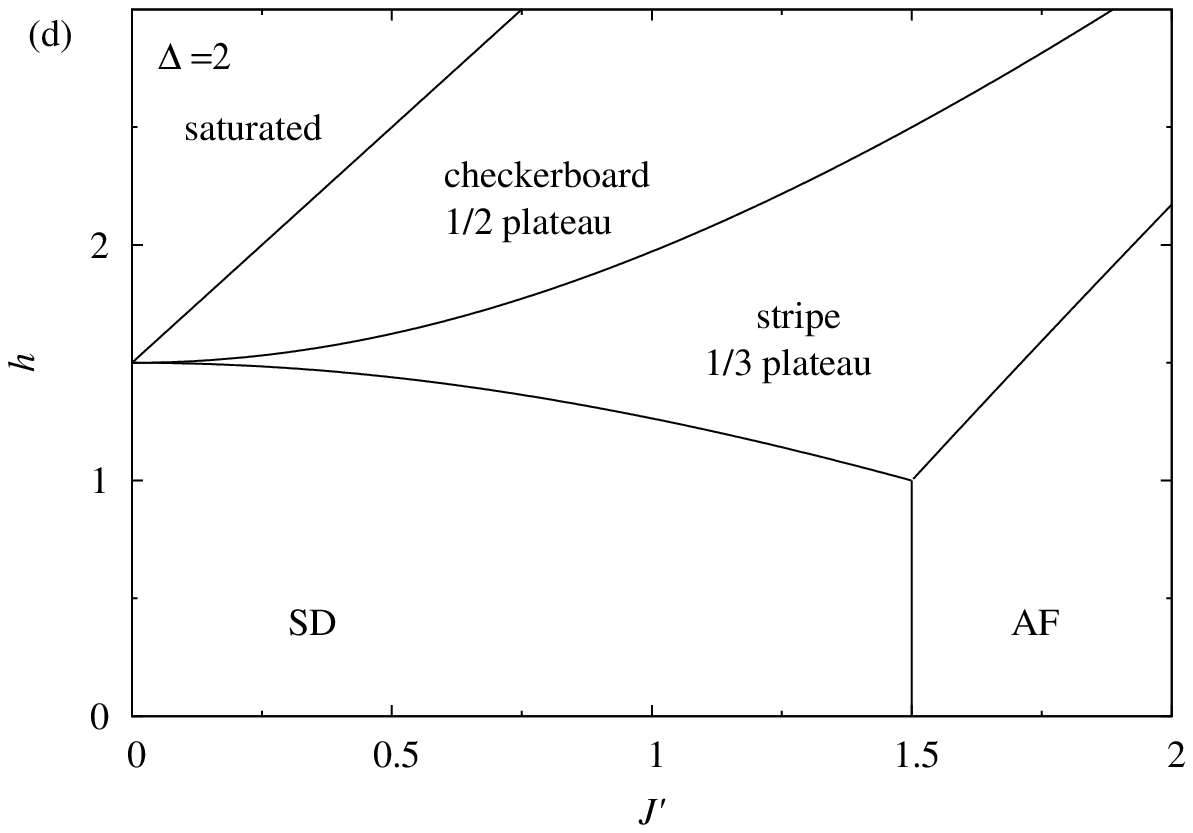, width=0.95\columnwidth}
\caption{Ground-state phase diagram of the spin-1/2 Ising-Heisenberg model on the Shastry-Sutherland model in the $J'-h$ plane for several values of the exchange anisotropy: (a) $\Delta=0.0$, (b) $\Delta=0.25$, (c) $\Delta=1.0$, (d) $\Delta=2.0$.}
\label{fig_gs}
\end{figure*}

Let us conclude our discussion about the ground state of the spin-1/2 Ising-Heisenberg model on the Shastry-Sutherland lattice by summarizing our findings. The zero-field ground state is either formed by the singlet-dimer phase for weaker Ising inter-dimer couplings \mbox{$J'<(1+\Delta)J/2$} or by the classical antiferromagnetic phase for stronger Ising inter-dimer couplings \mbox{$J'>(1+\Delta)J/2$}. The singlet-dimer phase remains the ground state at sufficiently small magnetic fields $h<h_1$ when \mbox{$J'<(1+\Delta)J/2$} and similarly the ground state remains in the antiferromagnetic phase at small enough magnetic fields $h<h_2$ when $J'>(1+\Delta)J/2$. The singlet-dimer and antiferromagnetic phases are replaced with the stripe 1/3-plateau phase, which becomes the ground state for intermediate magnetic fields \mbox{$h_3>h>h_1$} provided \mbox{$J'<(1+\Delta)J/2$} and, respectively, \mbox{$h_3>h>h_2$} if \mbox{$J'>(1+\Delta)J/2$}. The checkerboard 1/2-plateau phase is energetically favored over the stripe 1/3-plateau phase for magnetic fields $h>h_3$, and this ground state persists up to the saturation field $h<h_4$. For strong enough magnetic field $h>h_4$, the system ends up in the saturated paramagnetic phase with fully polarized dimers along the external magnetic field. 

The model with ferromagnetic Ising interaction $J'<0$ can be considered in the same manner as above. Taking $\gamma=1/2$ we find that in zero field the singlet state remains the ground state until $|J'|<(1+\Delta)J/2$, otherwise the ferromagnetic phase becomes favorable. In contrast to $J'>0$, there are no fractional plateaux. The magnetization jumps at 
\begin{eqnarray}
\label{h_fm}
h_c=\frac{(1+\Delta)J}{2}-|J'|
\end{eqnarray}
from zero in non-magnetic singlet-dimer phase to the maximal value in the saturated phase. To prove this statement, one has to use $\gamma$ determined from the condition $E(1,0,0)=E(0,1,0)$

To provide a more complete understanding of the overall ground-state behavior, we have plotted in Fig.~\ref{fig_gs} ground-state phase diagrams of the spin-1/2 Ising-Heisenberg model on the Shastry-Sutherland lattice for a few selected values of the exchange anisotropy $\Delta$. A comparison between the displayed ground-state phase diagrams allows us to clarify the effect of quantum fluctuations pertinent to the $XXZ$ Heisenberg intra-dimer interaction, the strength of which is controlled by the exchange anisotropy $\Delta$. It can be seen from Fig.~\ref{fig_gs}(a) that we have correctly recovered in the Ising limit $\Delta = 0$ the ground-state phase diagram of the spin-1/2 Ising model on the Shastry-Sutherland lattice reported previously by Dublenych [\onlinecite{dubl12}], which involves the Ising-dimer phase, the antiferromagnetic phase, the stripe 1/3-plateau phase and the saturated paramagnetic phase. It is noteworthy that all aforementioned phases become purely classical in the limiting case $\Delta=0$, i.e. there is no quantum reduction of local magnetizations within the stripe 1/3-plateau phase and no quantum entanglement between two-fold degenerate antiferromagnetic states within the Ising-dimer phase due to the complete lack of quantum fluctuations. On the other hand, the macroscopically degenerate Ising-dimer phase is transformed into the unique singlet-dimer phase with a perfect quantum entanglement between two antiferromagnetic states once the exchange anisotropy in the $XXZ$ Heisenberg intra-dimer coupling becomes non-zero (i.e. $\Delta>0$). It should be emphasized, moreover, that the ground-state phase diagram of the spin-1/2 Ising-Heisenberg model on the Shastry-Sutherland lattice remains qualitatively unchanged for any non-zero value of the exchange anisotropy $\Delta>0$ [c.f. Figs.~\ref{fig_gs}(b),(c),(d)]. The most fundamental difference between the spin-1/2 Ising and Ising-Heisenberg models on the Shastry-Sutherland lattice thus consists in the presence of the checkerboard 1/2-plateau phase in the ground-state phase diagram of the latter model, which is however totally absent in the ground-state phase diagram of the former model. A more subtle difference can be still found within the stripe 1/3-plateau phase even though this phase is present in the ground-state phase diagram of the Ising as well as Ising-Heisenberg model. In fact, the stripe 1/3-plateau phase undergoes according to Eq. (\ref{qrm}) a quantum reduction of local magnetizations of the Ising-Heisenberg model with $\Delta>0$ in contrast to fully saturated local magnetizations of the Ising model with $\Delta=0$. The quantum reduction of local magnetizations within the stripe 1/3-plateau state is the stronger, the greater the transversal part of the $XXZ$ Heisenberg intra-dimer interaction is (i.e. the greater the parameter $\Delta$ is). As far as two intermediate plateau states are concerned, one may generally observe the following general trends: (i) the easy-axis exchange anisotropy shrinks the width of the checkerboard 1/2-plateau until it completely disappears in the Ising limit $\Delta=0$; (ii) the easy-plane exchange anisotropy shrinks a width of the stripe 1/3-plateau although this plateau state does not entirely vanish in the $XX$ limit $\Delta \to \infty$.  
 
Next, let us make a few comments on the ground-state boundaries between different phases, where an extremely high macroscopic degeneracy may come into play. For instance, one may formulate an effective hard-core square model at the boundary between the saturated paramagnetic phase and the checkerboard 1/2-plateau phase, because the energies of three-dimer clusters with the configurations $(1,1,1)$, $(1,1,0)$, $(1,0,1)$ and $(0,1,0)$ must inevitably become equal. If starting from the fully saturated state we may place the singlet state on a dimer without cost of any energy. However, two singlets cannot be placed on nearest-neighbor dimers due to the restriction on the allowed configurations, which leads to an effective hard-core repulsion between particles representing singlets.

On the other hand, the ground state can be built up from any combination of the cluster configurations (0,0,0), (1,0,0) and (0,1,0), which are of equal energy at the boundary between the singlet-dimer phase and the stripe 1/3-plateau phase. Therefore, the triplon state can be created in the singlet-dimer phase with the special conditions of hard-core repulsion: two triplons cannot be placed on the nearest-neighbor dimers as well as on the next-nearest-neighbor dimers in a vertical (horizontal) direction for the horizontal (vertical) dimer, respectively. It is worth mentioning that the identical hard-core constraint for triplons was previously deduced for the spin-1/2 Heisenberg model on the Shastry-Sutherland lattice.\cite{mats13} The stripe 1/3-plateau phase can be thus viewed as the state with maximally dense packing of triplets, which still satisfies the afore-described hard-core constraint. Of course, other states with a lower density of triplets are also allowed by the hard-core constraint, whereas these states have the same energy as the singlet-dimer and the stripe 1/3-plateau phase at their ground-state boundary determined by the critical field $h_1$. This actually means that more complex ground states of the Heisenberg model such as 1/8-, 1/6- or 1/4-plateau states coexist together with the singlet-dimer and the stripe 1/3-plateau ground states along their ground-state phase boundary.\cite{mats13,taki13}

The boundary between the antiferromagnetic phase and the stripe 1/3-plateau phase is somewhat different. Namely, the three-dimer configuration (-1,1,0) can be additionally realized at the respective boundary besides the configurations (1,-1,1), (-1,1,-1) and (1,0,0), (0,1,0), which are building block of the antiferromagnetic and stripe 1/3-plateau phases. With regard to this, the boundary between these two ground states includes a lot of unexpected spin configurations. For instance, the ferromagnetic chain in the stripe 1/3-plateau phase can be extended to a set of ferromagnetically ordered chains, whereas the neighboring chains are magnetized in opposite directions with respect to each other and side chains are directed along the magnetic field. The situation at the boundary between the stripe 1/3-plateau phase and the checkerboard 1/2-plateau phase is quite similar. Any random spin configuration involving antiferromagnetic and ferromagnetic stripes is possible whenever the ferromagnetic stripes are separated from each other by one or two antiferromagnetic stripes.

Last but not least, let us compare our exact results for the ground state of the spin-1/2 Ising-Heisenberg model on the Shastry-Sutherland lattice with the known results for the analogous but fully quantum spin-1/2 Heisenberg model on the Shastry-Sutherland lattice obtained within the framework of various numerical approaches.\cite{koga00,weih02,corb13,jlou12,miya03,momo00,misg01,meng08,dori08,isae09,aben08,taki13,mats13,corb14} The zero-field ground states of the Ising-Heisenberg and Heisenberg models are quite similar in two limiting cases corresponding either to the weak inter-dimer coupling $J' \ll 1$ or to the strong inter-dimer coupling $J' \gg 1$. As a matter of fact, the singlet-dimer phase is the ground state of the Ising-Heisenberg as well as of Heisenberg model in the limiting case of weak inter-dimer coupling $J' \ll 1$, while the quantum reduction of local magnetization is the only relevant difference between the classical and quantum antiferromagnetic ground state of the Ising-Heisenberg and Heisenberg models in the other limiting case of the strong inter-dimer coupling $J' \gg 1$. Hence, the most substantial difference between the zero-field ground states of both these models can be detected at moderate values of the inter-dimer interaction $J' \approx 1$. It is worth recalling that the Ising-Heisenberg model with the isotropic Heisenberg intra-dimer interaction shows a direct first-order phase transition between the singlet-dimer phase and the antiferromagnetic phase at the specific value of the inter-dimer interaction $J'=1$ in contrast to the more complex behavior of the full quantum Heisenberg model, which exhibits an additional plaquette zero-field ground state in a range of moderate values of the inter-dimer coupling $0.675<J'<0.765$.\cite{koga00,weih02,miya03,corb13,jlou12} 

\begin{figure}[bt]
\epsfig{file=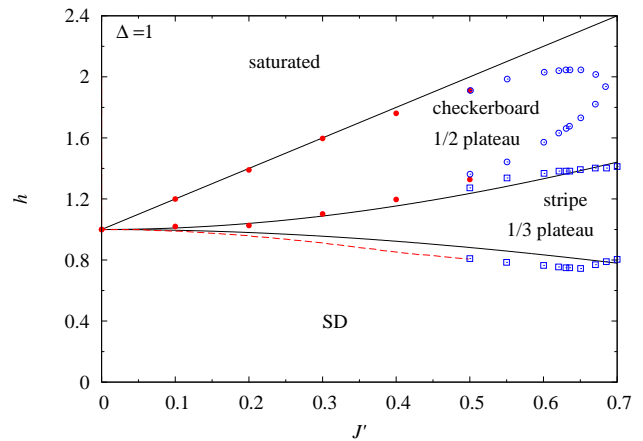, width=0.95\columnwidth}
\caption{(Color online) Comparison between the ground-state phase diagrams of the spin-1/2 Ising-Heisenberg and Heisenberg models on the Shastry-Sutherland model. Solid (black) lines represent the ground-state phase boundaries for the Ising-Heisenberg model (\ref{ham0}) with the isotropic Heisenberg intra-dimer coupling $\Delta=1$. Filled circles are exact diagonalization data of the pure quantum Heisenberg model for $N=36$ spins and the broken curve represents the CORE results for the same model adapted from Ref.~[\onlinecite{aben08}].
Empty squares and circles show the boundaries of the 1/3- and 1/2-plateaux obtained with iPEPS in Ref.~[\onlinecite{mats13}]. }
\label{fig_gs2}
\end{figure}

The ground-state phase diagram of the spin-1/2 Ising-Heisenberg model on the Shastry-Sutherland lattice in the $J'-h$ plane is confronted in Fig.~\ref{fig_gs2} with the analogous ground-state phase diagram of the spin-1/2 quantum Heisenberg model on the Shastry-Sutherland lattice adapted from the numerical data reported in Refs.~[\onlinecite{aben08,mats13}]. Although there is still some controversy about the microscopic nature, size and total number of magnetization plateaus of the spin-1/2 Heisenberg model on the Shastry-Sutherland lattice in a non-zero magnetic field, the microscopic nature of the stripe 1/3-plateau and checkerboard 1/2-plateau has been firmly corroborated by numerous precise numerical methods along with a few more subtle 1/4- and 2/5-plateaus the nature of which is nowadays under intensive debate\cite{miya03,momo00,misg01,aben08,meng08,dori08,isae09,taki13,mats13,corb14,folt14}. It is quite remarkable that the simplified Ising-Heisenberg model correctly reproduces the microscopic nature of the stripe 1/3-plateau and the checkerboard 1/2-plateau of the full quantum Heisenberg model, whereas it gives through the exact eigenvector (\ref{1o3-phase}) some additional insight into the microscopic origin of the stripe 1/3-plateau state and the related quantum reduction of local magnetizations. In addition, it surprisingly turns out that the ground-state phase diagrams of the Ising-Heisenberg and Heisenberg models are in a relatively good quantitative agreement not only in the limit of weakly interacting dimers $J' \to 0$ but up to moderate values of the inter-dimer interaction $J' \approx 0.5$. It is quite tempting to conjecture, moreover, that more subtle stripe 1/4- and 2/5-plateau states of the quantum Heisenberg model are merely stabilized by means of the transverse ($XX$) component of the inter-dimer interaction, because these states coexist in the Ising-Heisenberg model at the singlet-dimer vs. stripe 1/3-plateau phase boundary and respectively, at the stripe 1/3-plateau vs. the checkerboard 1/2-plateau phase boundary. This result would imply that the Ising-Heisenberg model can be considered as a good starting point for the perturbative treatment of the full quantum Heisenberg model. 

Let us complete this section by the discussion of the application to some real compounds. In SrCu$_2$(BO$_3$)$_2$, early thermodynamic measurements\cite{miya03} as well as the recent determination of the boundaries of the 1/3- and 1/2-plateaux\cite{seba08,mats13}  point to a ratio
$J'/J\approx 0.63$. For this ratio, our Ising-Heisenberg model shows quite close results for the boundaries of the 1/3-plateau, while the boundaries of the 1/2-plateau are quite different from those of the Heisenberg model (see Fig.~\ref{fig_gs2}). This discrepancy is caused by the wide region of 1/3- and 2/5-supersolid phases below 1/2-plateau and the spin-liquid-like phase above it that were observed in the Heisenberg model.\cite{mats13} Therefore, the upper boundary of 1/2-plateau is rather related to the saturation field of the corresponding Heisenberg model. We can briefly consider another compound with the magnetic structure the Shastry-Sutherland model, (CuCl)Ca$_2$Nb$_3$O$_{10}$.\cite{kode01,kage05,tsuj14} It corresponds to the isotropic Heisenberg model with strong antiferromagnetic intra-dimer and ferromagnetic inter-dimer couplings, and shows a transition from a spin-singlet ground state to the magnetized phase at the critical field $7.8$~T which correspond to a Zeeman energy of 11.1~K.\cite{tsuj14} From Eq.~(\ref{h_fm}) we can estimate the relation between intra-dimer and inter-dimer interactions as $J-|J'|\approx 11.1$~K.
The results for the Heisenberg and Ising-Heisenberg model perfectly coincide in this particular case, since they both describe a direct
field-induced transition from the phase of uncorrelated singlets to polarized dimers. The main reason for this surprising quantitative
agreement is that the quantum ($XY$) part of the inter-dimer interaction has no effect on both aforementioned phases of the corresponding
Heisenberg model.

\section{Conclusions}
\label{sec-conclusions}
The present work deals with the ground-state behavior of the spin-1/2 Ising-Heisenberg model on the Shastry-Sutherland lattice with the $XXZ$ Heisenberg intra-dimer interaction and the Ising inter-dimer interaction. Exact ground states of the model have been obtained by two independent procedures leading to equivalent effective Hamiltonians: the former one takes advantage of a local unitary transformation in order to establish a rigorous mapping correspondence with an effective classical spin-1 model, while the latter method leads to an effective hard-core boson model by a graph-based continuous unitary transformation. Apart from the exact ground states and ground-state phase diagrams we have also studied in some detail the degeneracy at particular phase boundaries.   

It has been demonstrated that the spin-1/2 Ising-Heisenberg model on the Shastry-Sutherland lattice exhibits a zero-temperature magnetization curve with just two intermediate plateaus at 1/3 and 1/2 of the saturation magnetization. The 1/3-plateau corresponds to a regular alternation of diagonal stripes of polarized dimers with two diagonal stripes of spin-singlet-like dimers, while a checkerboard ordering of singlets and polarized triplets takes place at the 1/2-plateau. The microscopic nature of the remarkable stripe 1/3-plateau has been thoroughly investigated with the help of the corresponding exact eigenvector, which shows that the quantum reduction of the local magnetization within this peculiar ground state is due to the competition between the Ising inter-dimer coupling and the transverse part of the $XXZ$ Heisenberg intra-dimer coupling. 

The rigorous results for the spin-1/2 Ising-Heisenberg model on the Shastry-Sutherland lattice have been also compared with the analogous results for the purely classical Ising and full quantum Heisenberg models. It has been verified that the ground-state phase diagrams of the Ising-Heisenberg and Heisenberg models are in a relatively good quantitative accordance up to moderate values of the inter-dimer coupling $J' \approx 0.5$. In addition, it has been shown that the 1/8-, 1/6-, and 1/4-plateaus coexist at the phase boundary between the singlet-dimer phase and the stripe 1/3-plateau phase and similarly, the stripe 2/5-plateau coexists at the phase boundary between the stripe 1/3-plateau and checkerboard 1/2-plateau. This result suggests that the exactly solved Ising-Heisenberg model could be used as a good starting point for a perturbative treatment of its full quantum Heisenberg counterpart model in order to find out how the transverse component of the $XXZ$ inter-dimer interaction can stabilize those plateau states. This issue is a challenging task left for future investigation. 

\begin{acknowledgments}
T.V. acknowledges the financial support provided by the National Scholarship Programme of the Slovak Republic for the Support of Mobility of Students, PhD Students, University Teachers, Researchers and Artists. J.S. acknowledges financial support provided by a grant from The Ministry of Education, Science, Research, and Sport of the Slovak Republic under Contract No. VEGA 1/0234/12, the ERDF EU (European Union European regional development fond) grant provided under the contract No. ITMS26220120005 (activity 3.2) and by grants of the Slovak Research and Development Agency under Contracts No. APVV-0132-11 and No. APVV-0097-12. F.M. acknowledges the financial support of the Swiss National Fund.

\end{acknowledgments}

\end{document}